\documentclass[11pt,twocolumn,tighten]{aastex63}

\usepackage{apjfonts}
\usepackage{url}
\usepackage{hyperref}
\usepackage{natbib}
\usepackage{amsmath,amstext,amssymb}
\usepackage[caption=false]{subfig} 
\usepackage{xcolor, fontawesome}
\usepackage{color}
\usepackage{enumitem}

\newcommand{\kms}{{km\,s$^{-1}$}}

\newcommand{\bprp}{G_{\rm BP} - G_{\rm RP}}

\newcommand{\caltech}{Department of Astronomy, MC 249-17, California Institute of Technology, Pasadena, CA 91125, USA}
\newcommand{\mitkavli}{MIT Kavli Institute and Department of Physics, 77 Massachusetts Avenue, Cambridge, MA 02139}

\newcommand{\princeton}{Department of Astrophysical Sciences, Princeton University, 4 Ivy Lane, Princeton, NJ 08540, USA}

\newcommand{\howard}{Department of Physics and Astronomy, Howard University, Washington DC, 20059}
\newcommand{\goddard}{Center for Research and Exploration in Space Science and Technology, and X-ray Astrophysics Laboratory, NASA/GSFC, Greenbelt, MD 20771, USA}

\newcommand{\nstarssearched}{65{,}760}
\newcommand{\nlcssearched}{180{,}017}

\newcommand{\nuniqdipflagged}{{368}} 
\newcommand{\nallcands}{66}
\newcommand{\ncqvsnodebunked}{63}
\newcommand{\ngoods}{50}
\newcommand{\nmaybes}{13}
\newcommand{\ndebunked}{3}

\newcommand{\nnotfieldbanyan}{61}
\newcommand{\ngoodhighruwe}{16}
\newcommand{\nmaybehighruwe}{0}
\newcommand{\ngoodweakruwe}{21}
\newcommand{\nmaybeweakruwe}{2}
\newcommand{\nbothdipfourier}{32}
\newcommand{\nyesdipnofourier}{23}
\newcommand{\nyesfouriernodip}{8}
\newcommand{\nrvscatterflag}{three}

\newcommand{\ngoodmultperiodflag}{22}
\newcommand{\nmaybemultperiodflag}{3}
\newcommand{\ngoodruweandmultperiod}{15}

\newcommand{\ngoodweakruweandmultperiod}{18}

\begin{document}

\title{Transient Corotating Clumps Around Adolescent Low-Mass Stars From Four Years of TESS}

\correspondingauthor{Luke G. Bouma}
\email{luke@astro.caltech.edu}

\received{2023 September 8}
\revised{2023 November 3}
\accepted{2023 November 6}
\shorttitle{Four Years of Complex Periodic Variables} 

\shortauthors{Bouma, Jayaraman, et al.}


\author[0000-0002-0514-5538]{Luke~G.~Bouma}
\altaffiliation{51 Pegasi b Fellow}
\affiliation{\caltech}

\author[0000-0002-7778-3117]{Rahul~Jayaraman}
\affiliation{\mitkavli}

\author{Saul~Rappaport}
\affiliation{\mitkavli}

\author[0000-0001-6381-515X]{Luisa M. Rebull}
\affiliation{\caltech}

\author{Lynne~A.~Hillenbrand}
\affiliation{\caltech}

\author[0000-0002-4265-047X]{Joshua N. Winn}
\affiliation{\princeton}

\author[0000-0003-4062-0776]{Alexandre David-Uraz}
\affiliation{\howard}
\affiliation{\goddard}

\author[0000-0001-7204-6727]{G\'asp\'ar \'A. Bakos}
\affiliation{\princeton}


\begin{abstract}
  Complex periodic variables (CPVs) are stars that exhibit highly
  structured and periodic optical light curves.  Previous studies have
  indicated that these stars are typically disk-free pre-main-sequence
  M dwarfs with rotation periods ranging from 0.2 to 2 days.  To
  advance our understanding of these enigmatic objects, we conducted a
  blind search using TESS 2-minute data of \nstarssearched\ K and M
  dwarfs with $T$$<$16\added{\,mag} and $d$$<$150\,pc.  We found \ngoods\
  high-quality CPVs, and subsequently determined that most are members
  of stellar associations.  Among the new discoveries are the
  brightest ($T$$\approx$9.5\added{\,mag}), closest ($d$$\approx$20\,pc), and
  oldest ($\approx$200\,Myr) CPVs known.  One exceptional object, LP
  12-502, exhibited up to eight flux dips per cycle. Some of these
  dips coexisted with slightly different periods, and the
  shortest-duration dips precisely matched the expected timescale for
  transiting small bodies at the corotation radius.  Broadly, our
  search confirms that CPVs are mostly young ($\lesssim$150\,Myr) and
  low-mass ($\lesssim$0.4\,$M_\odot$).  The flux dips characteristic
  of the class have lifetimes of $\approx$100 cycles, although stellar
  flares seem to induce sudden dip collapse once every few months.
  The most plausible explanation for these phenomena remains
  corotating concentrations of gas or dust.  The gas or dust is
  probably entrained by the star's magnetic field, and the sharp
  features could result from a multipolar field topology, a hypothesis
  supported by correspondences between the light curves of CPVs and of
  rapidly rotating B stars known to have multipolar magnetic fields.
\end{abstract}


\section{Introduction}
\label{sec:intro}

\begin{figure*}[!t]
	\begin{center}
		\subfloat{
			\includegraphics[width=0.8\textwidth]{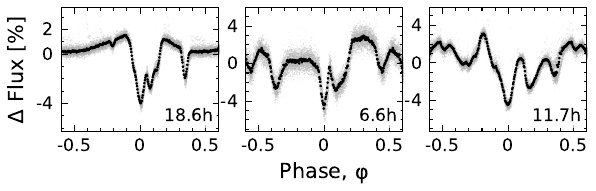}
		}
		
		\vspace{-0.62cm}
		\subfloat{
			\includegraphics[width=0.66\textwidth]{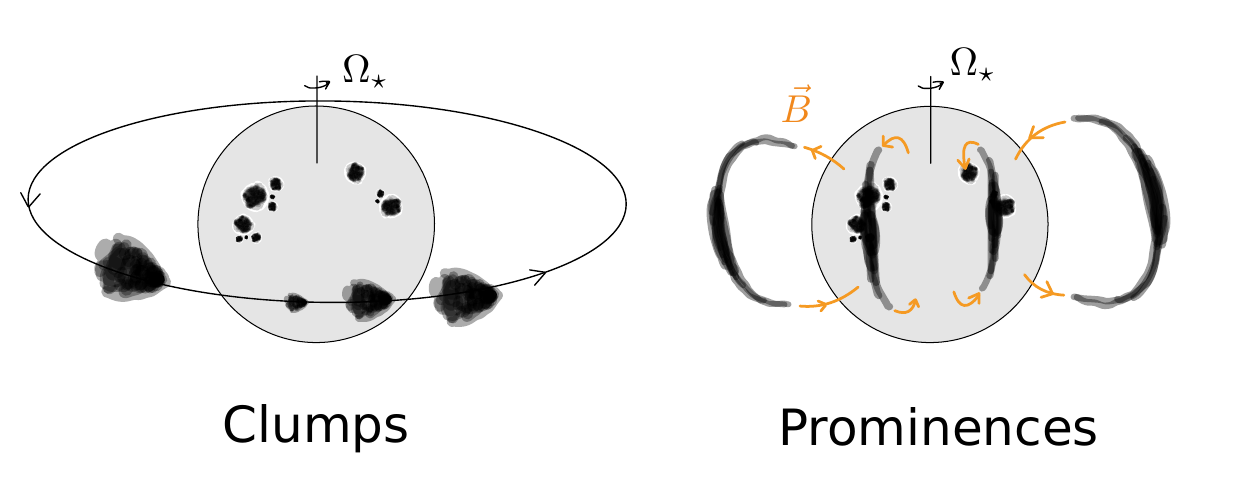}
		}
	\end{center}
	\vspace{-0.5cm}
	\caption{
		{\bf Complex periodic variables (CPVs)}:
		{\it Top:} Phase-folded TESS light curves for three CPVs.  Each
		panel shows the average of the data accumulated over one month,
		relative to the mean stellar brightness.
		Gray circles are raw 2-minute data; black circles are binned to
		300 points per cycle.  The period in hours is printed in the
		bottom right corner.  Left-to-right, the objects are LP 12-502
    (TIC~402980664; Sector~19), TIC~94088626 (Sector 10), and
    TIC~425933644 (Sector~28).
    {\it Bottom:} Cartoon configurations for magnetically-entrained
    corotating material.  The dust clump scenario (left) and gas
    prominence scenario (right) propose different opacity sources, and
    different occultation geometries.
	}
	\label{fig:f1}
\end{figure*}

All young stars vary in optical brightness, and the origin of such
variability is, in most cases, understood.  Well-explored sources of
optical variability include inhomogeneities on stellar surfaces such
as starspots and faculae \citep[e.g.][]{2021isma.book.....B},
occultations by circumstellar disks
\citep[e.g.][]{2017MNRAS.470..202B}, and, in geometrically favorable
circumstances, eclipses by stars and planets
\citep[e.g.][]{2020AJ....160...33R}.  More exotic sources of optical
variability that are potentially relevant to this work include
transiting exocomets (e.g.~$\beta$~Pic;
\citealt{2019A&A...625L..13Z}), disintegrating rocky bodies (e.g.
KOI-2700; \citealt{2014ApJ...784...40R}), and occultations by
circumstellar plasma clumps (e.g.~$\sigma$~Ori~E;
\citealt{2005ApJ...630L..81T,2005MNRAS.357..251T}).

Data from K2 \citep{2014PASP..126..398H} and TESS
\citep{2015JATIS...1a4003R} have revealed a new class of variable star
for which the root cause of variability is only beginning to become
clear: complex periodic variables (CPVs).  These objects are
identified from their optical light curves, which show nearly periodic
troughs that are either sharp or broad; these troughs are often
superposed on quasi-sinusoidal spot-like modulation
\citep{2017AJ....153..152S,2018AJ....155...63S,2019ApJ...876..127Z}.
Some CPVs show up to eight dips per cycle.  Most CPVs are
pre-main-sequence M dwarfs with ages of $\approx$5-150 million years
(Myr), and rotation periods of 0.2--2 days.  They are observed to
comprise $\approx$1-3\% of M dwarfs younger than 100 Myr
\citep{2016AJ....152..114R,2022AJ....163..144G}.  They generally do
not show near-infrared excesses indicative of dusty disks, but the
wavelength-dependent dip amplitudes of some CPVs is consistent with
reddening by dust
\citep{2017PASJ...69L...2O,2020AJ....160...86B,2022AJ....163..144G,2023MNRAS.518.2921K}.
The dip amplitudes and phases usually evolve gradually over tens to
hundreds of cycles, although they have occasionally been observed to
change abruptly within one cycle
\citep[e.g.][]{2017AJ....153..152S,2022ApJ...925...75P,2023ApJ...945..114P}.

The sharp features of CPV light curves can have durations as short as
5\% of the rotation period ($P_{\rm rot}$), which is too short to be
caused by starspots rotating into and out of view.  Starspots produce
flux variations with characteristic timescales of $P_{\rm rot}$ and
$0.5\,P_{\rm rot}$.  With finely-tuned viewing geometries, starspots
can produce dip durations as short as $\approx$$0.2\,P_{\rm rot}$, but
in such cases, limb darkening causes the dip amplitudes to be smaller
than the observed amplitudes of $\sim$1\% (see
\citealt{2017AJ....153..152S}, Figures 37-41).  Thus, a
``starspot-only'' scenario can be ruled out for many CPVs
\citep{2017AJ....153..152S,2019ApJ...876..127Z,2021MNRAS.500.1366K}.
Given that many CPVs cannot be explained by starspots alone, and
working under the assumption that all CPVs share the same basic
physical scenario, we discard the ``starspot-only'' model.
Instead, the correct explanation probably involves spatially
concentrated circumstellar material
\citep[e.g.][]{2017AJ....153..152S,2022AJ....163..144G}.

Figure~\ref{fig:f1} illustrates two proposed configurations for the extrinsic
material.  The first scenario invokes opaque dust ``clumps'' that
orbit near the Keplerian corotation radius [$R_{\rm c} =
(GM/\Omega^2)^{1/3}$, where $\Omega = 2\pi/P_{\rm rot}$] and
periodically transit the star
\citep[e.g.][]{2017AJ....153..152S,2017MNRAS.471L.145F,2023MNRAS.518.4734S}.
The second scenario invokes ``prominences'', long-lived condensations
of cool, dense, marginally-ionized gas that are embedded within the
hotter corona and that corotate with the star
\citep{1989MNRAS.238..657C,2019MNRAS.482.2853J,2022MNRAS.514.5465W}.
These hypothetical prominences are analogous to quiescent prominences
and filaments seen in the solar corona \citep[see
e.g.][]{2015ASSL..415.....V}, though rather than existing at a
fraction of the stellar radius as in the solar case, they would exist
at distances of a few stellar radii.  A final possibility is that an
optically-thick ring obscures a narrow band of the stellar photosphere
\citep{2019ApJ...876..127Z}; hot spots passing behind such a ring
could produce sudden dips.  We disfavor this scenario for reasons
described in Appendix~\ref{app:ring}.

While the dust clump and gas prominence hypotheses both invoke
magnetically-entrained material, the two pictures differ in the origin
and the composition of the occulting material.   ``Dust clumps''
invoke opacity from dust, which would need to be collisionally charged
\citep{2023MNRAS.518.4734S}, and which might be sourced from a
low-mass debris disk.  ``Gas prominences'' invoke opacity from
partially-ionized gas, perhaps bound-free transitions in hydrogen or a
molecular opacity.  The gas might be sourced from a stellar wind.
Unambiguous evidence in support of either scenario has yet to be
acquired.  Such evidence might include a spectroscopic detection of
silicate 10\,$\mu$m dust absorption during a dip, or perhaps detection
of transient Balmer-line excesses as a function of cycle phase,
similar to observations made in systems such as AB~Dor
\citep[see][]{1999ASPC..158..146C} or PTFO 8-8695
\citep{2016ApJ...830...15J}.

In both models, the corotation radius is the location at which matter
concentrates.  The empirical basis for this is that the sharp CPV
features are superposed over smooth, quasi-sinusoidal starspot
profiles.  The theoretical importance of the corotation radius has
been noted in previous studies of magnetic rotators
\citep[e.g.][]{1973ApJ...184..271L,1985Ap&SS.116..285N,1991ApJ...370L..39K,2005ApJ...634.1214L}.
In regions where the magnetic field dominates the flow (i.e.~$B^2/8\pi
> \rho v^2 /2$), matter is dragged along with the field lines.  Within
such regions, charged gas or dust can become trapped at corotation
because of the four relevant forces -- gravity, Lorentz, inertial
Coriolis, and inertial centrifugal -- the Lorentz and Coriolis only
act perpendicular to field lines, while gravity and the centrifugal
force are in balance at $R_{\rm c}$ \citep[e.g.][their
Section~2]{2005MNRAS.357..251T}.  Another way to phrase this statement
is that, in the corotating frame, the effective potential experienced
by charged particles tends to have local minima at $R_{\rm c}$; given
a flow from either the star or from a tenuous accretion disk, this
local potential minimum enables material to build up
\citep{2005MNRAS.357..251T}.

While theoretical heritage for understanding magnetic rotators exists,
CPVs have remained mysterious because they have been both hard to
discover and hard to characterize.   They have been hard to discover
because they are rare: CPVs comprise $\approx$1\% of the youngest
$\approx$1\% of M dwarfs \citep{2018AJ....155..196R}.  Out of the
millions of stars monitored by K2 and TESS, about 70 CPVs have been
reported to date
\citep{2016AJ....152..114R,2017AJ....153..152S,2018AJ....155...63S,2019ApJ...876..127Z,2020AJ....160...86B,2021AJ....161...60S,2022AJ....163..144G,2023ApJ...945..114P}.
They have been hard to characterize because many of the known CPVs are
faint; the initial K2 discoveries
\citep{2016AJ....152..114R,2017AJ....153..152S} were M2-M6 dwarfs at
distances $\gtrsim$100\,pc, with optical brightnesses of
$V$$\approx$15.5 to $V$$>$20.  At such magnitudes, high-resolution
time-series spectroscopy is out of reach with current facilities,
despite the potential utility of such observations.

In this work, we aim to find bright and nearby CPVs, since these
objects will be the most amenable to detailed photometric and
spectroscopic analyses.  To do this, we use 2-minute cadence data
acquired by TESS between 2018 July and 2022 September (Sectors 1-55;
Cycles 1-4).  We present our search methods in
Section~\ref{sec:methods}, and the resulting CPV catalog in
Section~\ref{sec:results}.  The observed evolution of many CPVs over a
two-year baseline is described in Section~\ref{sec:evoln}, including a
deep-dive into the behavior of an especially interesting object, LP
12-502.  We discuss a few implications in
Section~\ref{sec:discussion}, and conclude in
Section~\ref{sec:conclusion}.

Some comments on nomenclature are needed.  What we are calling
``complex periodic variables'' \citep{2023MNRAS.518.2921K} have also
been called ``complex rotators''
\citep{2019ApJ...876..127Z,2022AJ....163..144G,2023ApJ...945..114P},
``transient flux dips'', ``persistent flux dips'', and ``scallop
shells'' \citep{2017AJ....153..152S}.  The CPVs should not be
conflated with ``dippers'', which are classical T Tauri stars with
infrared excesses, and which show large-amplitude variability linked
to obscuring inner disk structures and accretion hot spots
\citep{2014AJ....147...82C,2021ApJ...908...16R}.  The phenomenology
and stellar properties of CPVs and dippers are quite different (though
see Sections~\ref{subsec:irexcess} and~\ref{subsec:discdippers}).  The
defining phenomenological features of the CPVs are that their light
curves are {\it complex}, relative to quasi-sinusoidal starspots, and
the complex features are {\it periodic}, meaning they typically repeat
for at least tens of days.  While rotation likely does play a central
role in explaining their physical behavior, the acronym for ``complex
rotator'' is already used in the astrophysical literature for cosmic
rays.  Given these considerations, we refer to the stars as complex
periodic variables (CPVs); our preferred explanation for their
behavior is that transient clumps of gas or dust orbit at their
corotation radii.

\section{Methods}
\label{sec:methods}

\subsection{Stellar selection function}
\label{subsec:selectionfn}

We searched for CPVs by analyzing the short-cadence data acquired by
TESS between 2018 July 25 and 2022 September 1 (Sectors 1-55).
Specifically, we used the 2-minute cadence light curves produced by
the Science Processing and Operations Center at the NASA Ames Research
Center \citep{2016SPIE.9913E..3EJ}.  While the TESS data products from
these sectors also included full frame images with cadences of 10 and
30 minutes for a larger number of sources, we restricted our attention
to the 2-minute data for the sake of uniformity and simplicity in data
handling.  In exchange, we sacrificed both completeness and
homogeneity of the selection function.  While TESS cumulatively
observed $\approx$90\% of the sky for at least one lunar month between
2018 July and 2022 September, the 2-minute cadence data were collected
for only a subset of observable stars that were preferentially nearby
and bright \citep[see][]{2021PASP..133i5002F}.  The total 2-minute
data volume from Sectors 1-55 included 1{,}087{,}475 short-cadence
light curves, which were available for 428{,}121 unique stars.


To simplify our search, we defined our target sample as stars with
2-minute cadence TESS light curves satisfying the following four
conditions:
\begin{align}
  T &< 16 \quad&(\mathrm{Amenable\ with\ TESS}) \label{eq:one}\\
  \bprp &> 1.5 \quad&(\mathrm{Red\ stars\ only})\\
  M_{\rm G} &> 4 \quad&(\mathrm{Dwarf\ stars\ only})\\
  d &< 150\,{\rm pc} \quad&(\mathrm{Close\ stars\ only}) \label{eq:four}.
\end{align}
Here, $M_{\rm G} = G + 5\log(\varpi_{\rm as}) + 5$ is the Gaia
$G$-band absolute magnitude, $\varpi_{\rm as}$ is the parallax in
units of arcseconds, and $d$ is a geometric distance defined by
inverting the parallax and ignoring any zero-point correction.  We
performed this selection by cross-matching TIC8.2
\citep{2019AJ....158..138S,2021arXiv210804778P} against the Gaia DR2
point-source catalog \citep{2018A&A...616A...1G}.  We opted for Gaia
DR2 rather than DR3 because the base catalog for TIC8 was Gaia DR2,
which facilitated a one-to-one crossmatch using the Gaia source
identifiers.  The target sample ultimately included \nstarssearched\ M
dwarfs and late-K dwarfs, down to $T$$<$16 and out to $d$$<$150\,pc.
For stars with multiple sectors of TESS data available, we searched
for CPV signals independently.  In total, our \nstarssearched\ star
target list included \nlcssearched\ month-long light curves.

\begin{figure}[!t]
	\begin{center}
		\centering
		\includegraphics[width=0.46\textwidth]{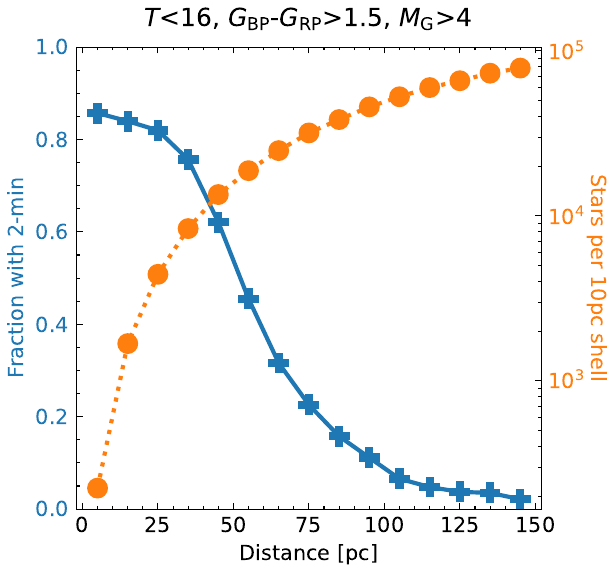}
		\vspace{-0.2cm}
		\caption{
      {\bf Completeness of the TESS 2-minute data for late-K and early
      M dwarfs near the Sun, from Sectors 1-55.}  The orange dotted
      curve shows the number of stars in successive radial shells,
      each with a width of 10\,pc.  To be part of our selection
      function, these stars must meet the following conditions: they
      must be red dwarf stars ($\bprp$>$1.5$; $M_{\rm G}$>4) amenable
      for TESS observations ($T$<16).  The blue solid curve shows the
      fraction of such stars with at least one sector of TESS 2-minute
      cadence data acquired between Sectors 1-55.
		}
			\vspace{-0.5cm}
		\label{fig:completeness}
	\end{center}
\end{figure}

We assessed the completeness of our selection function by comparing
the number of stars with TESS Sector 1-55 short-cadence data against
the number of Gaia DR2 point sources.  We required all stars to meet
conditions~\ref{eq:one}--\ref{eq:four}.  The results are shown in
Figure~\ref{fig:completeness}.  TESS 2-minute data exist for
$\approx$50\% of $T$$<$16 M and late-K dwarfs at $\approx$50\,pc.
Within 20\,pc, $\gtrsim$80\% of the $T$$<$16 M and late-K dwarfs have
at least one sector of short-cadence data.  Beyond 100\,pc,
$\lesssim$10\% of such stars have any short-cadence data available.
This can be translated into our sensitivity for the lowest mass stars
by considering that the spectral type of a $T$=16 star at $d$=50\,pc
is $\approx$M5.5V, corresponding to a main-sequence mass of
$\approx$0.12\,$M_\odot$.

\subsection{CPV discovery}
\label{subsec:discoverymethods}

Prior to this study, most CPVs have been found by visually examining
all the light curves of stars in young clusters
\citep{2016AJ....152..114R,2017AJ....153..152S,2023ApJ...945..114P},
or by flagging light curves with short periods and strong Fourier
harmonics for visual inspection \citep{2019ApJ...876..127Z}.  In this
work, we implemented a new search approach based on counting the
number of sharp local minima in phase-folded light curves, while also
using the Fourier approach.  We applied these two search techniques
independently to our \nstarssearched\ targets.

\subsubsection{Counting dips}
\label{subsec:counting}

The dip counting technique aims to count sharp local minima in
phase-folded light curves.  The most remarkable CPVs often show three
or more dips per cycle, which distinguishes them from other types of
variables such as synchronized and spotted binaries (RS CVn stars).

For our dip-counting pipeline, we began with the {\tt PDC\_SAP} light
curves for each sector \citep{2017ksci.rept....8S}, removed non-zero
quality flags, and normalized the light curve by dividing out its
median value.  We then flattened the light curve using a 5-day sliding
median filter, as implemented in \texttt{wotan}
\citep{2019AJ....158..143H}.  We computed a periodogram of the
resulting cleaned and flattened light curve, opting for the
\citet{1978ApJ...224..953S} phase dispersion minimization (PDM)
algorithm implemented in \texttt{astrobase}
\citep{2021zndo...1011188B} due to its shape agnosticism.  If a period
$P$ below 2 days was identified, we reran the periodogram at a finer
grid to improve the accuracy of the period determination.

Once a star's period was identified, we binned the phased light curve
to 100 points per cycle.  To separate sharp local minima from smooth
spot-induced variability, we then iteratively fit robust penalized
splines to the wrapped phase-folded light curve, excluding points more
than two standard deviations away from the local continuum
\citep{2019AJ....158..143H}.  The wrapping procedure is discussed
below.  In this fitting framework, the maximum number of equidistant
spline knots per cycle is the parameter that controlled the meaning of
``sharp'' --- we allowed at most 10 such knots per cycle, though for
most stars fewer knots were preferred based on cross-validation using
an $\ell^2$-norm penalty.  An example fit is shown in panel {\it (e)}
of Figure~\ref{fig:vet}.

We then identified local minima in the resulting residual light curve
using the SciPy \texttt{find\_peaks} utility
\citep{2020NatMe..17..261V}, which is based on comparing adjacent
values in an array.  For a peak to be flagged as significant, we
required it to have a width of at least $0.02\,P$, and a height of at
least twice the noise level.  The noise level was defined as the
68$^{\rm th}$ percentile of the distribution of the residuals from the
median value of $\delta f_i \equiv f_i - f_{i+1}$, where $f$ is the
flux and $i$ is an index over time.  In panel {\it (e)} of
Figure~\ref{fig:vet}, automatically-identified local minima are shown
with the gray triangles.

Wrapping is necessary to eliminate edge effects when fitting the light
curve and when identifying local minima in the residuals.  A phased
light curve would usually cover phases $\phi \in [ 0,1 ]$.  We instead
performed the analysis described above using a phase-folded light
curve spanning $\phi \in [-1,2 ]$, which was created by duplicating
and concatenating the ordinary phase-folded light curve.  The free
parameters we adopted throughout the analysis -- for instance the
maximum number of spline knots per cycle, and the height and depth
criteria for dips -- were chosen during testing based on the desire to
correctly re-identify a large fraction ($>$90\%) of previously known
CPVs, while also being able to consistently reject common false
positives such as spot-induced variability and eclipsing binaries.

In short, CPV candidates were identified by requiring a peak PDM
period below two days and the presence of at least three sharp local
minima, based on at least one sector of the TESS 2-minute data.
Candidates were then inspected visually as described in
Section~\ref{subsec:visual}.

\subsubsection{Fourier analysis}
\label{subsec:fourier}

We performed an independent search using a Fourier-based approach,
following \citet{2019ApJ...876..127Z} and \citet[][their
Section~1.3]{2023MNRAS.524.4220P}.  Starting with the {\tt PDC\_SAP}
light curves, we normalized each light curve, and then re-binned it
into equal width 2-minute bins to account for the uneven spacing in
the TESS data, as well as the data gap caused by satellite downlink
during each sector.  We then padded the data to ensure that the light
curve had a length that was a power of two, as described by
\citeauthor{2019ApJ...876..127Z}\ After taking the Fourier transform
of the padded light curve using {\tt numpy.fft}, we searched for peaks
with a significance exceeding 12-$\sigma$ within a set of 500
frequency bins.

Peaks of significance were found for $\approx$10\% of the searched
stars.  For all such cases, we generated an interim ``summary sheet''
with information about the star, its full and folded light curves,
Fourier transform, potential contaminating stars, and information
about these contaminating stars.  We then reviewed each summary sheet,
and tentatively classified each light curve based on visual inspection
of its morphology (with common categories including eclipsing binary,
CPV, RS CVn, and cataclysmic variable).

\subsubsection{Manual vetting}
\label{subsec:visual}

\begin{figure*}[!tp]
	\begin{center}
		\centering
		\includegraphics[width=0.99\textwidth]{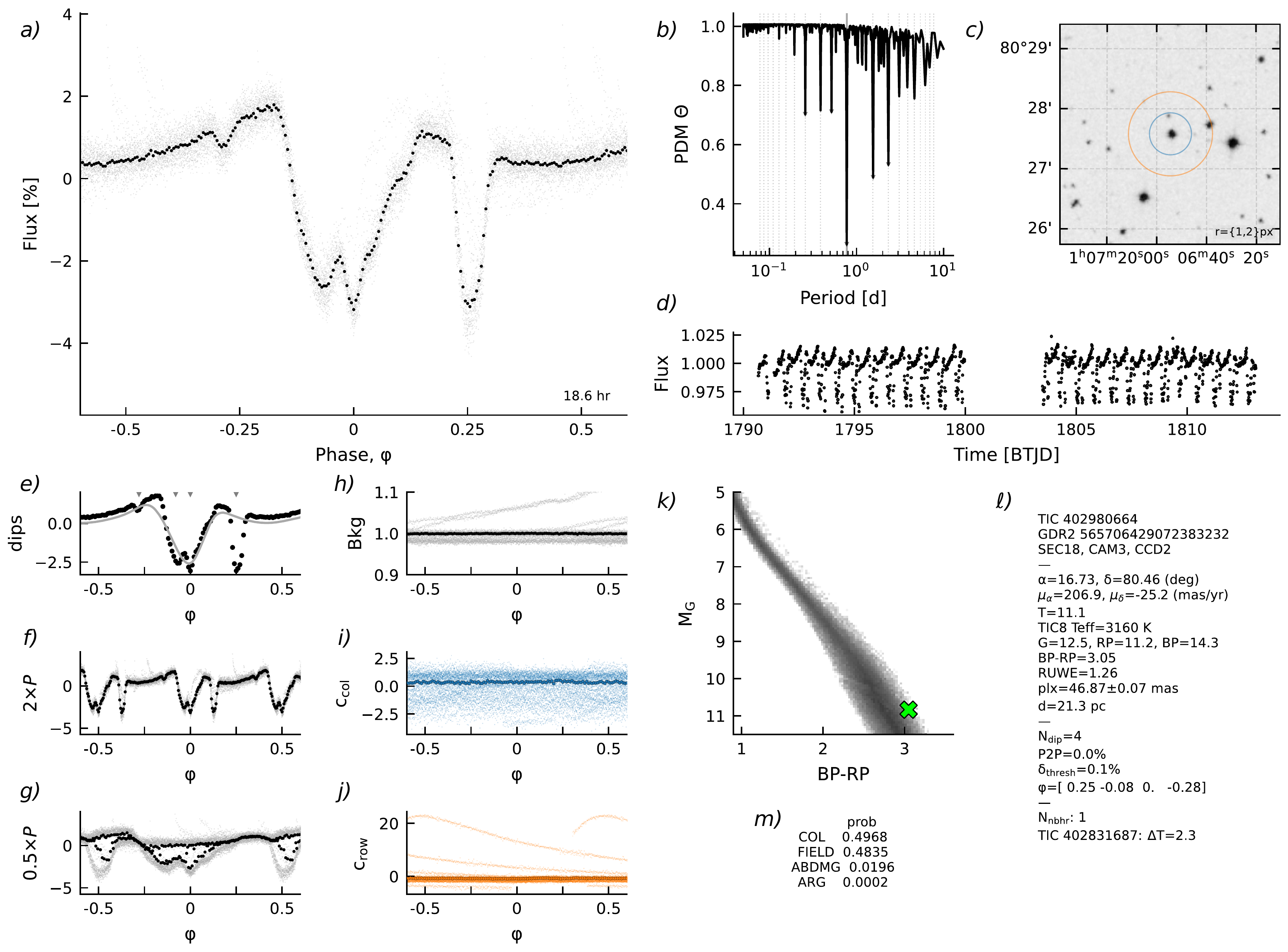}
		\vspace{-0.3cm}
		\caption{
      {\bf Validation plots used to classify CPVs}.  The complete
      figure set contains one image
      per sector for each of the \nallcands\ objects in
      Table~\ref{tab:thetable}, and
      is accessible both through the online journal and via
      \url{https://zenodo.org/record/8327508}. Panels are as follows.
			{\it a)}: Phase-folded light curve; gray points are raw 2-minute
			data and black points are binned to 200 points per cycle.
      The adopted period is given in the lower-right corner.
			{\it b)}: Phase-dispersion minimization (PDM) periodogram.
			Dotted lines show up to the 10$^{\rm th}$ harmonic and
			subharmonic.
      {\it c)}: DSS finder chart, with 21$''$ and 42$''$ radius
      circles for scale.  One TESS pixel has a full side length of
      21$''$.
			{\it d)}: Cleaned light curve, binned to 20-minute cadence, in
			Barycentric TESS Julian Date (BTJD).
			{\it e)}: Phase-folded light curve, binned to 100 points per
			cycle.  The gray line denotes the spline-fit to the
			wrapped phase-folded light curve, and small gray triangles
			denote automatically identified local minima.
			{\it f)}: Phase-folded light curve at twice the peak period.
			{\it g)}: Phase-folded light curve at half the peak period.
			{\it h)}: Phase-folded time-series within the ``background''
			aperture defined in the SPOC light curves.
			{\it i)}: Phase-folded flux-weighted centroid in the column
			direction.
			{\it j)}: Phase-folded flux-weighted centroid in the row
			direction.
      {\it k)}: Gaia DR2 color--absolute magnitude diagram. The gray
      background denotes stars within 100\,pc from
      \citet{2021A&A...649A...6G}.
      {\it l)}: Information from Gaia DR2, TIC8, and the automated
      dip-counting search pipeline.  ``Neighbors'', abbreviated
      ``nbhr'', are listed within apparent distances of 2 TESS pixels
      if $\Delta T$$<$2.5.
      {\it m)}: BANYAN\,$\Sigma$ v1.2 association probabilities,
      calculated using positions, proper motions, and the parallax.
		}
		\label{fig:vet}
	\end{center}
\end{figure*}

We homogeneously assessed whether the objects identified using the
dip-counting (Section~\ref{subsec:counting}) and Fourier
(Section~\ref{subsec:fourier}) approaches were consistent with
expectations for CPVs by assembling the data shown in
Figure~\ref{fig:vet}.  We labeled a star as a ``good'' CPV if it met
all of the following criteria for at least one TESS sector:
\vspace{-2pt}
\begin{itemize}[leftmargin=*]
  \setlength\itemsep{-2pt}
  \item $P<2$\,days.
  \item At least three dips per cycle, or else otherwise oddly-shaped
    dips relative to expectations for quasi-sinusoidal
    starspot-induced modulation.
  \item Persistent dips over multiple consecutive rotation cycles.
\end{itemize}
\vspace{-2pt}
We also noted \replaced{''possible'' CPVs for which the classification
was more ambiguous, and}{a few stars with potentially oddly-shaped dips
as ``ambiguous'' CPVs, and a few interesting} ``false
positives'' that are definitely not CPVs.  The most common false
positives for both the Fourier and dip-counting techniques were
eclipsing binaries, ordinary spotted rapid rotators, and light curves
that were complex due to multiple stars contributing to the
photometric aperture.  Our specialized dip-counting pipeline flagged
\nuniqdipflagged\ unique stars for visual inspection; about 20\% were
subsequently labeled either good or \replaced{possible}{ambiguous} CPVs.  From the more
general Fourier pipeline, $\approx$0.5\% of stars that passed the
12-$\sigma$ peak threshold were eventually classified as CPVs.

\subsection{Stellar properties}
\label{subsec:starprops}

\subsubsection{Ages}
While most of our target stars were field stars, color--absolute
magnitude diagrams suggested that most CPVs tended to be on the
pre-main-sequence (e.g.\ panel {\it (k)} of Figure~\ref{fig:vet}).  We
therefore estimated stellar ages by checking for probabilistic spatial
and kinematic associations between the CPVs and known clusters in the
solar neighborhood.  For most stars in our sample, we did this using
BANYAN\,$\Sigma$
\citep{2018ApJ...856...23G}.\footnote{\url{https://github.com/jgagneastro/banyan_sigma},
git commit \texttt{394b486}} This algorithm calculates the probability
that a given star belongs to either the field, or to any of 27 young
clusters (``associations'') within 150\,pc of the Sun.  This is
achieved by modeling the field and cluster populations as multivariate
Gaussian distributions in 3-D position and 3-D velocity space.  We
used the Gaia DR2 sky positions, proper motions, and distances to
calculate the membership probabilities.  BANYAN\,$\Sigma$ in turn
analytically marginalizes over the radial velocity dimension.  The
probabilities returned by this procedure are qualitatively helpful,
but should be interpreted with caution because the assumption of
Gaussian distributions is questionable for most groups within the
solar neighborhood \citep[see e.g.][Figure~10]{2021ApJ...917...23K}.

For a few cases where BANYAN\,$\Sigma$ yielded ambiguous results, we
consulted the meta-catalog of young, age-dated, and age-dateable stars
\replaced{provided}{assembled} by \citet{2022AJ....163..121B}, and also searched the local
volume around each star for co-moving
companions.\footnote{\url{https://github.com/adamkraus/Comove}, git
commit \texttt{278b372}\added{; see also \citet{2021AJ....161..171T}.}}\added{ A few important sources in the former meta-catalog included the
Theia groups from \citet{KounkelCovey2019} and \citet{2020AJ....160..279K}, and the SPYGLASS stars from Table~1 of \citet{2021ApJ...917...23K}.} Finally, to provide a base for comparison, we
also ran the BANYAN\,$\Sigma$ membership analysis on our entire
\nstarssearched\ target star sample.

\subsubsection{Effective temperatures, radii, and masses}

We determined the stellar effective temperature and radii for the CPVs
by fitting the broadband spectral energy distributions (SEDs); we then
estimated the masses by interpolating against the sizes, temperatures,
and ages of the PARSEC v1.2S models
\citep{2012MNRAS.427..127B,2014MNRAS.444.2525C}.

For the SED fitting, we used \texttt{astroARIADNE}
\citep{2022MNRAS.513.2719V}.  We adopted the BT-Settl stellar
atmosphere models \citep{Allard2012} assuming the
\citet{2009ARA&A..47..481A} solar abundances, and the
\citet{2006MNRAS.368.1087B} water line lists.  The broadband
magnitudes we considered included $GG_{\rm BP}G_{\rm RP}$ from Gaia
DR2, $Vgri$ from APASS, $JHK_{\rm S}$ from 2MASS, SDSS $riz$, and the
WISE $W1$ and $W2$ passbands.  We omitted UV flux measurements from
our SED fit to avoid any possible bias induced by chromospheric UV
excess.  We omitted WISE bands $W3$ and $W4$ due to reliability
concerns.  \texttt{astroARIADNE} compares the measured broadband flux
measurements against pre-computed model grids, and by default fits for
six parameters: $\{ T_{\rm eff}, R_\star, A_{\rm V}, \log g, [{\rm
Fe/H}], d \}$.  The distance  prior is drawn from
\citet{2021AJ....161..147B}.  The surface gravity and metallicity are
generally unconstrained.  Given our selection criteria for the stars,
we assumed the following priors for the temperature, stellar size, and
extinction:
\begin{align}
  T_{\rm eff} / {\rm K}    &\sim \mathcal{U}(2000, 8000), \\
  R_\star / R_\odot  &\sim \mathcal{U}(0.1, 1.5), \\
  A_{\rm V} / {\rm mag}    &\sim \mathcal{U}(0, 0.2),
\end{align}
for \deleted{$\mathcal{N}$ the Gaussian and }$\mathcal{U}$ the uniform
distribution\deleted{s, and $\mathcal{T}_{\rm N}(\mu, \sigma, a, b)$ a
truncated normal distribution with mean $\mu$, standard deviation
$\sigma$, and lower and upper bounds $a$ and $b$}.  We validated our
chosen upper bound on $A_{\rm V}$ using a 2MASS color-color diagram.
Finally, using \texttt{Dynesty} \citep{2020MNRAS.493.3132S}, we
sampled the posterior probability assuming the default Gaussian
likelihood, and set a stopping threshold of ${\rm d}\log \mathcal{Z} <
0.01$, where $\mathcal{Z}$ denotes the evidence.

With the effective temperatures and stellar radii from the SED fit, we
estimated the stellar masses by interpolating against the PARSEC
isochrones \citep[v1.2S;][]{2014MNRAS.444.2525C}.  The need for models
that incorporate some form of correction for \deleted{young, active }M dwarfs is
well-documented \citep[e.g.][]{2012ApJ...757..112B,2012ApJ...756...47S,2015ApJ...804..146D,2016A&A...593A..99F,2018AJ....155..225K,2019MNRAS.489.2615M,2020ApJ...891...29S}.
Plausible explanations for the disagreement between observed and
theoretical M dwarf colors and sizes include starspot coverage
\citep[e.g.][]{2017ApJ...836..200G} and \deleted{potentially }incomplete line
lists \citep[e.g.][]{2013A&A...556A..15R}.  In the PARSEC models,
\citet{2014MNRAS.444.2525C} performed an empirical correction to the
temperature--opacity relation drawn from the BT-Settl model
atmospheres, in order to match observed masses and radii of young
eclipsing binaries.  This is sufficient for our goal of estimating
stellar masses.  Given our estimates of $\{ \tilde{T}_{\rm eff},
\tilde{R}_\star, \tilde{t} \}$, and approximating their uncertainties
as Gaussian $\sigma_{\tilde{T}_{\rm eff}}$, $\sigma_{\tilde{R}_\star}$
and $\sigma_{\tilde{t}}$, we define a distance metric $\Delta$ to each
model PARSEC grid-point $\{ T_{\rm eff}, R_\star, t \}$ via
\begin{equation}
  \Delta^2 = 
  \left( \frac{\tilde{T}_{\rm eff} - T_{\rm eff}}{\sigma_{\tilde{T}_{\rm eff}}} \right)^2
  +
  \left( \frac{\tilde{R}_{\star} - R_{\star}}{\sigma_{\tilde{R}_{\star}}} \right)^2
  +
  \left( \frac{\tilde{t} - t}{\sigma_{\tilde{t}}} \right)^2,
\end{equation}
where the division by the uncertainties helps to assign equal
importance to each dimension.  The mass reported in
Table~\ref{tab:thetable} is the model mass that minimizes the
distance.  The reported uncertainties in the masses are based on
propagating the statistical uncertainties in the radii, temperatures,
and ages.

\subsubsection{Binarity}

The main types of binaries of interest in this work are those that are
unresolved, because they can lead to misinterpretations of the data.
For instance, unresolved binaries might produce multiple photometric
signals and hinder our ability to correctly identify the star hosting
the CPV signal.  Unresolved binaries could also bias photometric
magnitude and color measurements, which would affect our stellar
parameter estimates.  To attempt to identify binaries, we considered
the following lines of information.

{\it Radial velocity scatter}---We examined diagrams of the Gaia DR3
``radial velocity error'' as a function of stellar color for all
\ncqvsnodebunked\ CPVs and candidate CPVs.  Since this quantity
represents the standard deviation of the non-published Gaia RV time
series, outliers can suggest single-lined spectroscopic binarity\added{ \citep[e.g.][]{2022arXiv220611275C}}.  These plots showed two
clusters of stars, at $\lesssim$10\,\kms and
20-25\,\kms.  We therefore adopted a threshold of $20$\,km\,s$^{-1}$
to flag possible single-lined spectroscopic binaries, which selected
\nrvscatterflag\ stars: TIC~405910546, TIC~224283342, and
TIC~280945693.

{\it RUWE}---We examined plots of Gaia DR3 RUWE as a function of
color.\footnote{For an explanation of the renormalized unit weight
error (RUWE), see the GAIA DPAC technical note
\url{http://www.rssd.esa.int/doc_fetch.php?  id=3757412}.}  Elevated
RUWEs imply excess astrometric noise relative to a single-source
model.  This can be caused by marginally resolved binaries,  intrinsic
photometric variability, or intrinsic astrometric motion\added{
\citep[e.g.][]{2021AJ....162..128W}}.  Based on this exercise, we
adopted a threshold of RUWE$_{\rm DR3}>2$ to flag
sources with excess astrometric noise.  This threshold was met by
\ngoodhighruwe/\ngoods\ high-quality CPVs and by
\nmaybehighruwe/\nmaybes\ of the ambiguous CPVs.  The choice of the
threshold RUWE is somewhat subjective, since the RUWE distribution has
an extended tail \citep[e.g.][]{2022MNRAS.513.5270P}.  If we had
instead required RUWE$_{\rm DR3}>1.4$, \ngoodweakruwe/\ngoods\
high-quality CPVs and \nmaybeweakruwe/\nmaybes\ of the ambiguous
sample would have been flagged.

{\it Gaia DR3 non-single stars}---Gaia DR3 included a {\tt
non\_single\_star} column that flagged eclipsing, astrometric, and
spectroscopic binaries.  None of the stars in our CPV sample were
identified as \deleted{possible }binaries in this column.

{\it Multiple periodic TESS signals}---During our visual analysis of
the TESS light curves and PDM periodograms, we flagged sources with
beating light curves, and with PDM periodograms that showed multiple
periods.  For such cases, we then subtracted the mean CPV signal over
each sector, and repeated the phase-dispersion minimization analysis.
The resulting secondary periods, $P_{\rm sec}$, are listed in
Table~\ref{tab:thetable}; we required these to be at least 5\%
different from the primary period.  The majority of secondary signals
showed morphologies corresponding to starspot modulation.  This
process yielded \ngoodmultperiodflag/\ngoods\ high-quality CPVs with
secondary periods; \nmaybemultperiodflag/\nmaybes\ of the ambiguous
sample met the same criterion.  Of the \ngoodhighruwe\ good CPVs with
${\rm RUWE}_{\rm DR3}> 2$, \ngoodruweandmultperiod\ also showed
secondary periods in the TESS light curves.  Considering the weaker
threshold of ${\rm RUWE}_{\rm DR3}> 1.4$,
\ngoodweakruweandmultperiod/\ngoodweakruwe\ such CPVs showed secondary
TESS periods.  The latter results strongly suggest that the secondary
periods are associated with bound binary companions.

Table~\ref{tab:thetable} summarizes each of the sources of binarity
information into a single bitwise column.  We describe detailed
results concerning binarity in Section~\ref{subsec:resultsbinarity},
and summarize those results Section~\ref{subsec:discbinary}.

\begin{figure*}[!tp]
	\begin{center}
		\centering
		\includegraphics[width=0.98\textwidth]{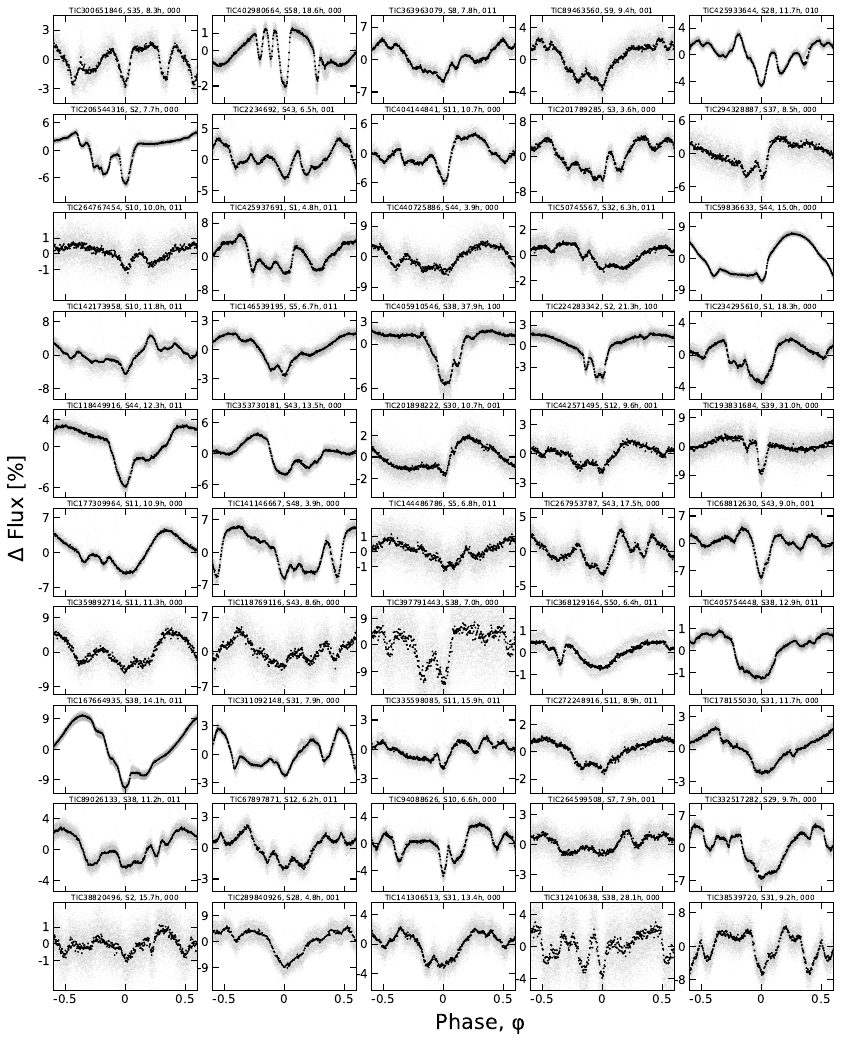}
    \vspace{-0.3cm}
		\caption{
      {\bf CPVs found in the TESS 2-minute data.}
      Phased TESS light curves over one month are shown for \ngoods\
      CPVs in the high quality sample.  Gray are raw 2-minute data;
      black bins to 300 points per cycle.  Objects are ordered such
      that sources with the most TESS data available are on top (see
      Section~\ref{sec:catalog}).  Zero phase is chosen to correspond
      to minimum light.  Each panel is labeled by the TIC identifier,
      the TESS sector number, the period in hours, and the three-bit
      binarity flag from Table~\ref{tab:thetable}, which denotes Gaia
      DR3 \texttt{radial\_velocity\_error} outliers (bit 1), Gaia DR3
      \texttt{ruwe} outliers (bit 2), and stars with secondary TESS
      periods (bit 3). 
		}
		\label{fig:cpvs}
	\end{center}
\end{figure*}

\section{Results}
\label{sec:results}

\subsection{CPV catalog}
\label{sec:catalog}

Table~\ref{tab:thetable} lists the \nallcands\ objects identified by
our search.  The \ngoods\ stars in the ``good'' sample demonstrated
what we deemed to be the key characteristics of the CPV phenomenon in
at least one TESS sector.  The classification of \nmaybes\ CPV
candidates was ambiguous, and the \ndebunked\ remaining objects were
notable false positives that we discuss below.  The \texttt{quality}
column in the table divides the three classes; additional data from
TESS or other instruments could help resolve the classification of the
ambiguous cases.   Of the \ncqvsnodebunked\ CPVs and candidate CPVs,
\nbothdipfourier\ were found using both the dip-counting and Fourier
techniques, \nyesdipnofourier\ were found using only the dip-counting
technique, and \nyesfouriernodip\ were found using only the Fourier
technique.  In the following, we will focus our discussion on the good
sample, irrespective of discovery method.  We will often refer to
stars by their TIC identifiers; these can be referenced against the
figures in most digital document readers using a ``find'' ({\tt
Ctrl+F}) utility.

Figure~\ref{fig:cpvs} is a mosaic of phased light curves for the
\ngoods\ CPVs.  The objects are sorted first in order of the number of
TESS 2-minute cadence sectors in which they clearly demonstrated the
CPV phenomenon, and secondarily by descending brightness.  The top
five objects by this metric are TIC~300651846 (12 sectors);
TIC~402980664 (7 sectors); TIC~89463560 (5 sectors); TIC~363963079 (5
sectors); and TIC~294328887 (4 sectors).  The brightest five CPVs span
9.3$<$$T$$<$11.1; the faintest five span 14.5$<$$T$$<$15.0.  The
fastest five have periods spanning 3.6\,hr$<$$P$$<$6.2\,hr, and the
slowest five span 27\,hr$<$$P$$<$38\,hr.

The light curves show between two and eight local minima per cycle.
Some stars show ordinary sinusoidal modulation during one portion of
the phased light curve, and highly structured modulation in the
remainder of the cycle (e.g. TIC~206544316, TIC~224283342,
TIC~402980664).  Others show structured modulation over the entire
span of a cycle (e.g. TIC~2234692, TIC~425933644, TIC~142173958).
Others show some mix between these two modes.

A small number of objects at first glance seem reminiscent of
eclipsing binaries, such as TIC~193831684, TIC~59836633, or
TIC~5714469.  We believe these cases are unlikely to be eclipsing
binaries due to the additional coherent peaks and troughs in the light
curves, which are distinct from any binary phenomena of which we are
aware.

\subsection{Ages of CPVs}
\label{subsec:ageresults}

Of our \ncqvsnodebunked\ confirmed and candidate CPVs,
\nnotfieldbanyan\ were associated with a nearby moving group or open
cluster, primarily using BANYAN\,$\Sigma$ as described in
Section~\ref{subsec:starprops}.\footnote{Two of the \nnotfieldbanyan\
memberships were made with low confidence and are flagged in
Table~\ref{tab:thetable}.  The assignment of TIC~397791443 to IC\,2602
was based not on BANYAN\,$\Sigma$ but instead on a literature search
\citep[e.g.][]{2020A&A...633A..99C}.}  The relevant groups are listed
in Table~\ref{tab:thetable}; their ages span $\approx$5-200\,Myr.  For
comparison, BANYAN\,$\Sigma$ assigned high-probability ($>$95\%) field
membership to 59{,}361 out of the \nstarssearched\ target stars.  Most
stars in our target sample are old; the CPVs returned by our blind
search are young.

The groups that contain the largest number of CPVs in our catalog are
Sco-Cen, Tuc-Hor, and Columba.  Six CPVs were also identified in the
Argus association \citep{2019ApJ...870...27Z}, which serves as an
indirect line of evidence supporting the reality and youth of that
group.  The large contribution from Sco-Cen is not surprising since
Sco-Cen contains the majority of pre-main-sequence stars in the solar
neighborhood, and many of its stars were selected for TESS 2-minute
cadence observations by guest investigators.  Given the
$\lesssim$$10\%$ completeness of \replaced{TESS}{our data} beyond 100\,pc
(Figure~\ref{fig:completeness}), there may be many more CPVs in
Sco-Cen that remain to be discovered.

There were two stars for which neither BANYAN\,$\Sigma$ nor a
literature search led to a confident association with any young group.
Both stars display CPV signals over multiple TESS sectors. Both are
photometrically elevated relative to the main sequence, an indication
of youth.  Both were also noted by \citet{2021ApJ...917...23K} as
being in the ``diffuse'' population of $<$50\,Myr stars near the Sun.

Our search confirms that the CPV phenomenon persists for at least
$\approx$150\,Myr.  Table~\ref{tab:thetable} includes three
$\approx$150\,Myr CPVs in AB~Dor \citep{2015MNRAS.454..593B}, a
$\approx$112\,Myr old Pleiades CPV \citep{2015ApJ...813..108D}, and a
similarly-aged Psc-Eri member \citep{2020A&A...639A..64R}.  To our
knowledge, TIC~332517282 in AB~Dor ($t$$=$$149^{+51}_{-19}$\,Myr;
\citealt{2015MNRAS.454..593B}) was the previous record-holder for the
oldest-known CPV \citep{2019ApJ...876..127Z,2022AJ....163..144G}; at
least one unambiguous CPV (EPIC~211070495) and a few other candidates
were also previously known in the Pleiades
\citep{2016AJ....152..114R}.  

The maximum age of CPVs might even exceed 200\,Myr, based on the
candidate membership of TIC~294328887 in the Carina Near moving group
\citep{2006ApJ...649L.115Z}.  The estimated age of this group, $200
\pm 50$\,Myr, is based on the lithium sequence of its G-dwarfs
\citep{2006ApJ...649L.115Z}, which shows a coeval population of stars
older than the Pleiades and younger than the 400\,Myr Ursa Major
moving group.  However, the formal BANYAN\,$\Sigma$ membership
probability is somewhat low (only 6\%), perhaps due to the missing
radial velocity.  This lack of information could be rectified by
acquiring even a medium-resolution spectrum.  An independent
assessment of the group's kinematics using Gaia data, and its rotation
sequence using TESS, could also bear on the question of whether
TIC~294328887 is a member.

\subsection{Infrared excesses of CPVs}
\label{subsec:irexcess}

Most CPVs in our catalog did not show infrared excesses in the
$W1$-$W4$ bands, which is typical for this class of object
\citep{2017AJ....153..152S}.  \replaced{Visually inspecting}{Inspecting} the SEDs of our
\nallcands\ star sample and the WISE images available through IRSA, we
labeled two objects as having reliable infrared excesses (both $W3$
and $W4$ fluxes are more than 3$\sigma$ above the photospheric
prediction): TIC~193136669 (TWA~34) and TIC~57830249 (TWA~33).
However, neither is considered a ``good'' CPV for the reasons that
follow.

Both of the stars with IR excesses are in the TW Hydrae association
($\approx$10\,Myr).  They have periods of 38\,hr and 44\,hr,
respectively.  In our initial labeling, we labeled both as
``ambiguous'' CPVs because the dips in their Sector~36 light curves
seemed to stochastically evolve over only one or a few cycles, which
is atypical for CPVs; their periods were also long in comparison with
most of the other CPVs.  Inspection of additional sectors clarified
that both sources are dippers, not CPVs (see the online plots in
Figure~\ref{fig:vet}).  For TIC~57830249, the Sector~10 light curve
shows completely different behavior from Sector~36, with variability
amplitudes of $\pm 50\%$ and no obvious periodicity.  TIC~57830249
also shows continuum emission at 1.3\,mm \citep{2015A&A...582L...5R},
which suggests that cold dust grains are present.

The dipper classification of TIC~193136669 is less obvious; the main
indication that it is a dipper is that Sectors 62 and 63 show its dips
appearing and disappearing within the span of one cycle.  None of the
CPVs in our sample exhibit this property.  Independently,
TIC~193136669 is known to have a cold disk of dust and molecular gas,
based on 1.3\,mm continuum emission and resolved $^{12}{\rm CO}(2-1)$
emission \citep{2015A&A...582L...5R}.  It was labeled a dipper by
\citet{2022ApJS..263...14C}; we agree with their designation, and
label it an ``impostor'' CPV in Table~\ref{tab:thetable}.
Section~\ref{subsec:discdippers} highlights plausible evolutionary
connections between CPVs and dippers in light of these
``misclassifications''.

\begin{figure*}[!t]
	\begin{center}
		\centering
		\includegraphics[width=\textwidth]{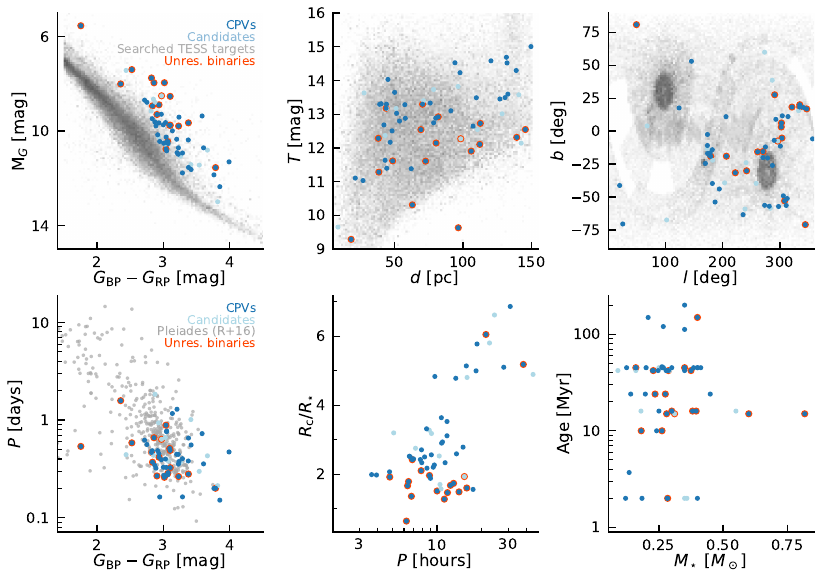}
		\vspace{-0.3cm}
		\caption{
			{\bf Properties of CPVs identified by our search}.
      CPVs are mostly pre-main-sequence M dwarfs, younger than
      $\approx$150 Myr, with rotation periods faster than
      $\approx$1\,day.  The \ngoods\ bona fide CPVs in
      Table~\ref{tab:thetable} are the dark blue dots; \nmaybes\
      ambiguous CPV candidates are light blue dots.  Unresolved
      binaries (red rings) are objects for which the Gaia DR3 radial
      velocity scatter exceeded 20\,\kms, or if Gaia ${\rm RUWE}_{\rm
      DR3}$$>$2 and multiple photometric signals were present in the
      TESS light curve.  The top panels show the \nstarssearched\
      target stars with 2-minute cadence TESS data as the shaded gray
      background; darker regions correspond to a larger relative
      number of searched stars.  The lower-left panel compares the
      rotation--color distribution of CPVs against the rotation
      periods of K and M dwarfs in the Pleiades from
      \citet{2016AJ....152..114R}.  The lower-middle panel plots the
      derived corotation radii $R_{\rm c} = (GM/\Omega^2)^{1/3}$ in
      units of stellar radii against the measured CPV periods, in
      units of hours.  Ages in the final panel are known from cluster
      membership.
		}
		\label{fig:catalogscatter}
	\end{center}
\end{figure*}

\subsection{Binarity of CPVs}
\label{subsec:resultsbinarity}

\subsubsection{Binary statistics}

A significant fraction of the CPVs show indications of unresolved
binarity.  Excess noise above the Gaia single-source astrometric model
is common (\ngoodhighruwe/\ngoods\ high-quality CPVs have RUWE$_{\rm
DR3}$$>$2), as is the presence of multiple periods in the TESS light
curves (\ngoodmultperiodflag/\ngoods).  Elevated astrometric noise
almost always implies multiple detectable TESS periods
(\ngoodruweandmultperiod/\ngoodhighruwe\ high-quality CPVs).  The
latter observation corroborates the claim that most sources with
RUWE$_{\rm DR3}$$>$2 are binaries with projected apparent separations
below 1$''$, and projected physical separations $\lesssim$50\,AU.
These observations are also in agreement with previous analyses of
multi-periodic low-mass objects discovered by K2, which found that
such systems are almost always binaries
\citep{2018AJ....156..138T,2018AJ....156..275S}.

\subsubsection{Do K dwarf CPVs exist?}
\label{subsec:massive}

To date, the only stars reported to show the CPV phenomenon are M
dwarfs, with typical stellar masses $\lesssim$0.3\,$M_\odot$
\citep{2017AJ....153..152S,2022AJ....163..144G}.  However the two most
massive CPVs in our sample, TIC~405754448 and TIC~405910546, were
assigned masses of $\approx$0.82\,$M_\odot$ and
$\approx$0.60\,$M_\odot$ respectively.  The next-highest mass in our
sample belongs to TIC~59836633 ($\approx$\replaced{0.48}{0.45}\,$M_\odot$), with all
remaining CPVs having masses $\lesssim$0.40\,$M_\odot$.

The locations of TIC~405754448 and TIC~405910546 in color--absolute
magnitude diagrams, combined with their probable membership in Lower
Centaurus Crux, support the conclusion that these stars have
relatively high masses.  However in detail, both objects are subject
to ambiguities in interpretation.  The TIC~405910546 light curve has a
unique shape, suggestive of an eclipsing binary.  Independently,
TIC~405910546 was one of only \nrvscatterflag\ CPVs flagged with a
Gaia DR3 radial velocity scatter exceeding 20\,\kms.  Combined, these
factors suggest that TIC~405910546 could be a pre-main-sequence
eclipsing binary; it should be studied further to clarify this
classification.

For the other object, TIC~405754448, the evidence for binarity is
stronger.  The RUWE$_{\rm DR3}$ statistic is 6.8, and the raw light
curves in Sectors 11, 37, and 38 show both the CPV signal with period
12.9\,hr and amplitude $\approx$1\% and an additional sinusoidal
signal with a period $\approx$6.5\,days and amplitude $\approx$0.3\%,
likely from a second star.  If TIC~405754448 is a K+M binary, then the
flux ratio between the primary and secondary would be expected to be
$\approx$10:1.  Thus, if the K star were the source of the CPV signal,
its intrinsic variability amplitude would be $\approx$1\%, while if
the M star were responsible its intrinsic variability amplitude would
be $\approx$10\%.

In short, these two objects suggest that the CPV phenomenon may extend
up in mass to pre-main-sequence K dwarfs, but more data are needed to
substantiate this claim.

\subsubsection{An astrophysical CPV false positive: TIC~435903839}

We originally classified TIC~435903839, with RUWE$_{\rm DR3}$=17.7, as
an ``ambiguous'' CPV with a 10.8\,hr period, because this period
minimized the dispersion in the phase-folded light curve.  More
careful inspection revealed an impostor: this source is a photometric
blend of two ordinary rotating stars with $P_0$=3.60\,hr, and
$P_1$=5.41\,hr, giving a beat period $(P_0^{-1} - P_1^{-1})^{-1}$ of
10.8\,hr.  This is a novel false positive scenario for CPVs: two rapid
rotators near the 3:2 period commensurability.  The beat between the
two rotation signals produces the apparent CPV signal.  Such false
positives can be excluded through careful accounting of all peaks in a
periodogram.  For instance, TIC~435903839 shows a peak at 16.27\,hr,
which is not an integer multiple of the dispersion-minimizing
10.82\,hr period.

\subsubsection{Multiple CPVs in the same system: TIC~425937691 and TIC~142173958}
\label{subsec:multis}

TIC~142173958 and TIC~425937691 both show evidence for two separate
CPV signals in their TESS light curves.  For TIC~142173958, the
signals have periods of 11.76\,hr and 12.84\,hr.  For TIC~425937691,
the two periods are 4.82\,hr and 3.22\,hr, near the 3:2 period
commensurability.  Given that both sources have two photometric
signals and elevated RUWEs, each source is probably an unresolved
binary consisting of two CPVs.  To our knowledge, these are the third
and fourth such systems known: EPIC~204060981 has two CPVs with
periods of 9.59\,hr and 9.12\,hr \citep{2018AJ....155...63S}, and
TIC~242407571 has two CPVs with periods of 11.33\,hr and 13.63\,hr,
near the 6:5 period commensurability \citep{2021AJ....161...60S}.


\section{Evolution of CPV Behavior}
\label{sec:evoln}

\begin{figure*}[!tp]
	\begin{center}
		\centering
		\includegraphics[width=\textwidth]{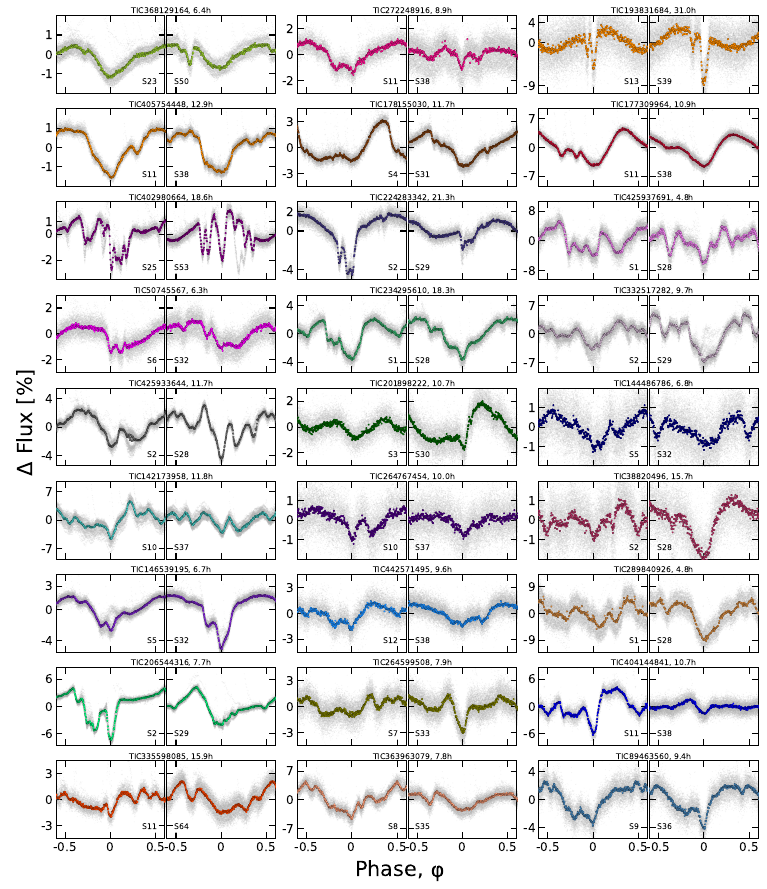}
		\vspace{-0.6cm}
		\caption{
			{\bf Evolution of CPV light curves over two years.}
			Out of the \ngoods\ CPVs in Figure~\ref{fig:cpvs}, 32 had
			2-minute cadence TESS data available for a baseline of at
			least two years; the 27 brightest are shown here due to space
			constraints.  Each panel shows one sector of TESS data, and is
			phased to its deepest minimum in flux.  Each panel's title shows
			the TIC identifier and period in hours.  Text insets
			show the TESS sector numbers, which generally span two years, or
			at least 1{,}000 cycles.  The vertical scale is fixed across
			sectors to clarify shape changes.  Gray circles are raw 2-minute
			data; colored circles bin to 300 points per cycle. 
		}
		\label{fig:evoln}
	\end{center}
\end{figure*}

\begin{figure*}[!tp]
	\begin{center}
		\centering
		\includegraphics[width=0.98\textwidth]{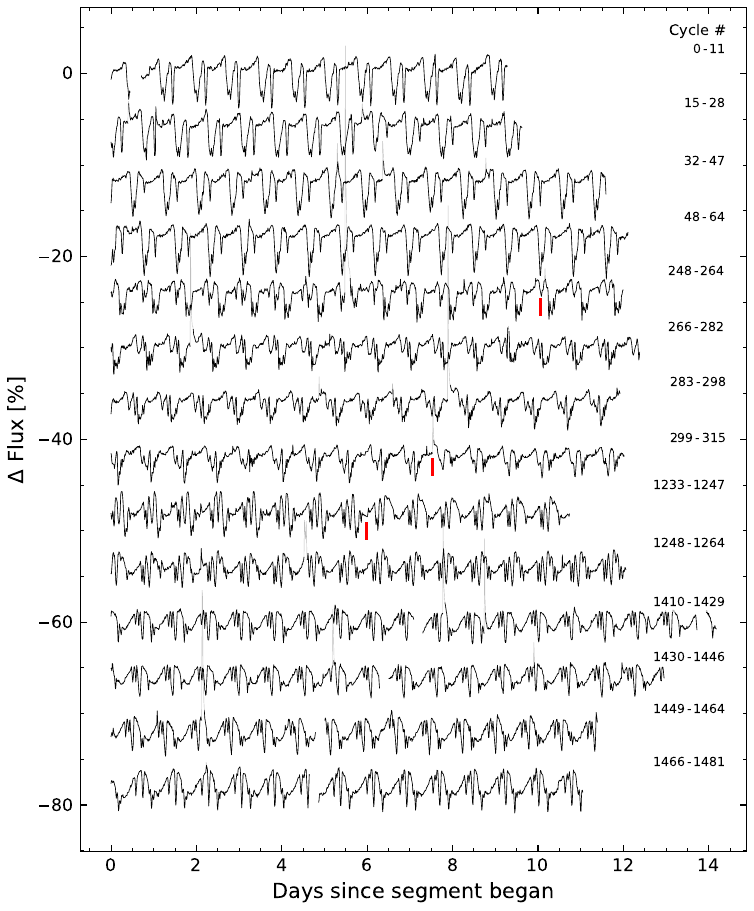}
		\vspace{-0.3cm}
		\caption{
      {\bf LP 12-502 (TIC~402980664) light curve}, where each time
      segment represents one TESS orbit.  Data were acquired in
      Sectors 18-19, 25-26, 53, and 57-58.  Flares are drawn in gray.
      The light curve is binned to 15-minute intervals so that there
      are 96 points per day, and each point is connected by a line.
      Data gaps have nothing plotted.  The red vertical lines
      highlight apparently instantaneous state changes in the shape of
      the dip pattern.  
		}
		\label{fig:lplc}
	\end{center}
\end{figure*}

\subsection{Evolution over two year baseline}
\label{subsec:twoyear}

Figure~\ref{fig:evoln} shows ``before'' and ``after'' views of 27 CPVs
for which TESS 2-minute cadence observations were available at least
two years apart.  Such a baseline was available for 32 of the \ngoods\
confirmed CPVs in our catalog; for plotting purposes we show the
brightest 27.  We have defined $\phi=0$ for each sector to be the time
of minimum light observed in that sector, rather than using a
consistent phase definition across multiple sectors.  This is because
for most of the sources we do not know the period at the precision
necessary to be able to accurately propagate an ephemeris over two
years.  The achievable period precision, $\sigma_P$, can be estimated
as
\begin{equation}
  \sigma_{P} = \frac { \sigma_{\phi} P } { N_{\rm baseline} },
  \label{eq:periodprecision}
\end{equation}
for $N_{\rm baseline}$ the number of cycles in the observed baseline
and $\sigma_{\phi}$ the phase precision with which any one feature
(e.g.~a dip, or the overall shape of the sinusoidal envelope) can be
tracked.  Assuming $\sigma_\phi$$\approx$$0.02$ and a 20-day baseline
over a single TESS sector yields
$\sigma_{P}$$\approx$$0.25^{+0.38}_{-0.14}$\,minutes for the
population shown in Figure~\ref{fig:evoln}; propagated forward 1{,}000
cycles yields a typical ephemeris uncertainty range of 2 to 11\,hours.
Measuring the period independently for each sector did not reveal
evidence for significant ($>$3$\sigma$) changes in period, implying a
period stability of $\lesssim$0.1\% over two years.

A few objects in Figure~\ref{fig:evoln} show the CPV phenomenon in one
sector, and only marginal signs or no sign of CPV behavior in the
other sector.  In our subjective assessment, cases for which at least
one sector would be flagged as ``ambiguous'' include
TIC~368129164 (Sector 23 might be labeled an EB),
TIC~177309964 (Sector 38 would be simply a rotating star),
TIC~404144841 (Sector 38 looks like a rotating star),
TIC~201898222 (Sector 3 looks like a rotating star),
TIC~144486786 (Sector 32 might be an RS CVn),
and
TIC~38820496 (Sector 28 might be an RS CVn).
TIC~193831684, assessed on a single-sector basis, would probably be
labeled an eclipsing binary---in fact, \citet{2021ApJ...912..123J}
already gave this source such a label.  However, based on the shape
evolution between Sectors 13 and 39, it is a CPV.

Based on the fraction of sources that ``turned off'', the
observed shape evolution implies that CPVs have an on-off duty cycle
of $\approx$75\%.  Correcting for the duty cycle might be important in
population-level estimates of the intrinsic frequency of the CPV
phenomenon \citep[e.g.][]{2022AJ....163..144G}.

\begin{figure*}[!t]
	\begin{center}
		\centering
		\includegraphics[width=1\textwidth]{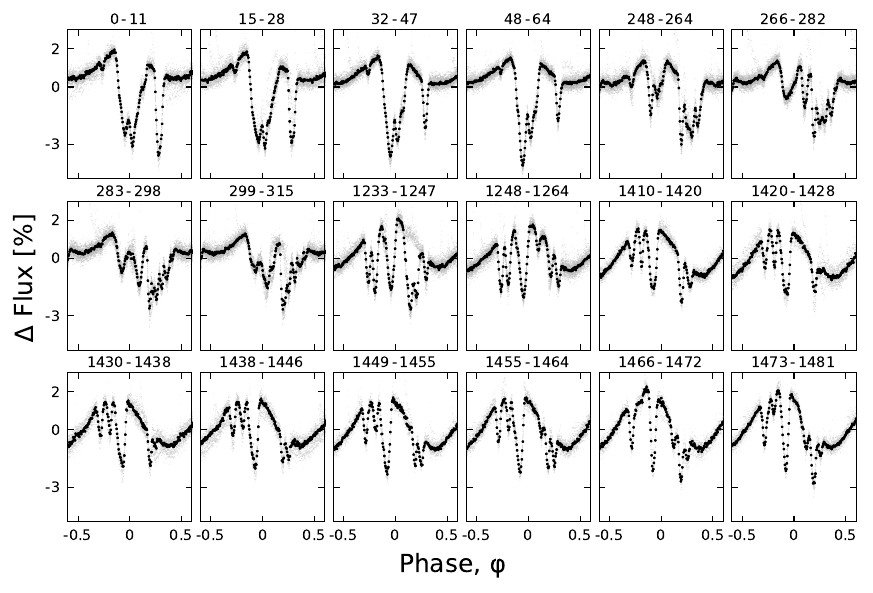}
		\vspace{-0.7cm}
		\caption{
			{\bf Evolution of LP 12-502} ($P$=18.5611\,h) at fixed period and
      epoch over three years.  Each panel shows one (averaged) TESS
			orbit; small text denotes relative cycle number.  There are 200
			binned black points per cycle.  The TESS pointing law dictates
			the large time gaps between cycles 64-248, 315-1233, and
			1264-1410; larger gaps tend to yield larger shape changes.  The
			dips usually evolve over tens to hundreds of cycles.  However
			cycles 1233-1264 show a dip that switched from a depth and
			duration of 3\% and 3\,hr to 0.3\% and 1\,hr over less than one
			cycle (cf.~Figure~\ref{fig:lplc}).
		}
		\label{fig:lp}
	\end{center}
\end{figure*}


\begin{figure*}[!t]
	\begin{center}
		\subfloat{
			\includegraphics[width=0.49\textwidth]{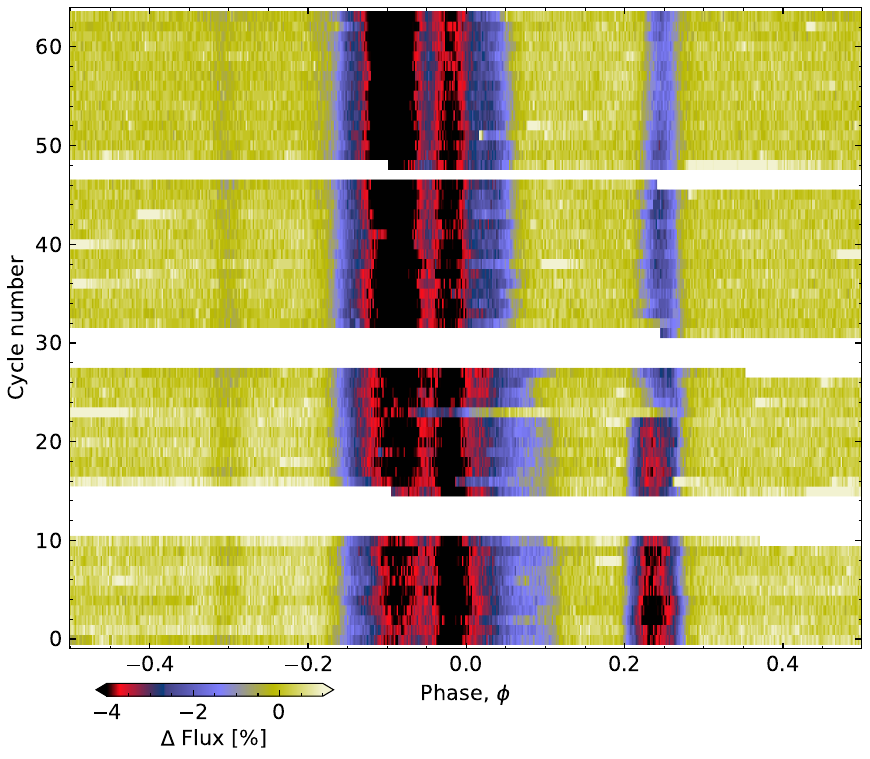}
			\includegraphics[width=0.49\textwidth]{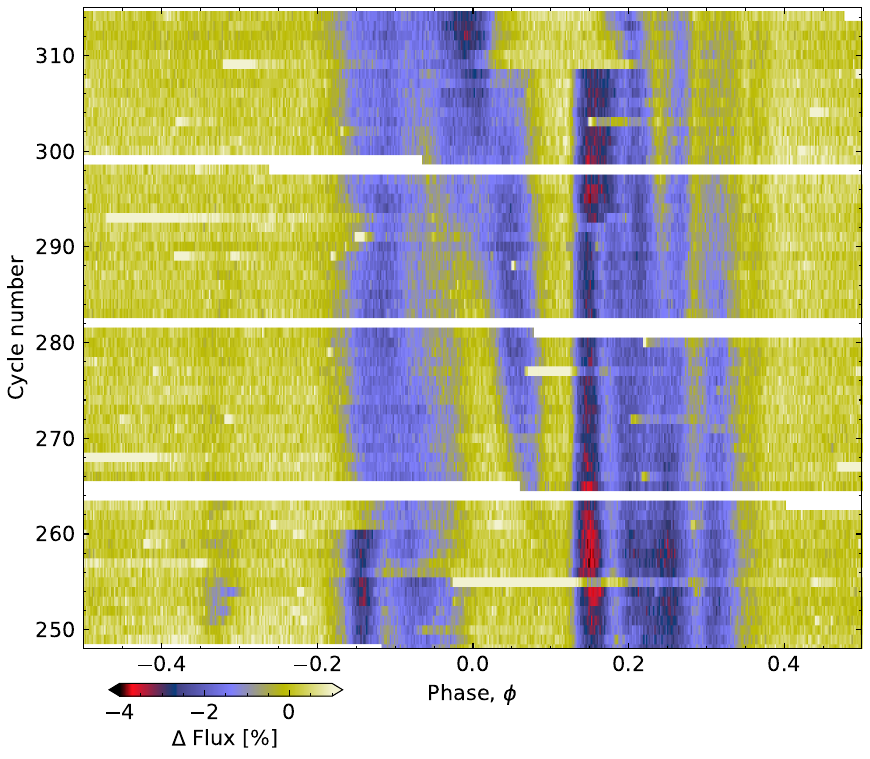}
		}
		\vspace{-0.2cm}
		
		\subfloat{
			\includegraphics[width=0.49\textwidth]{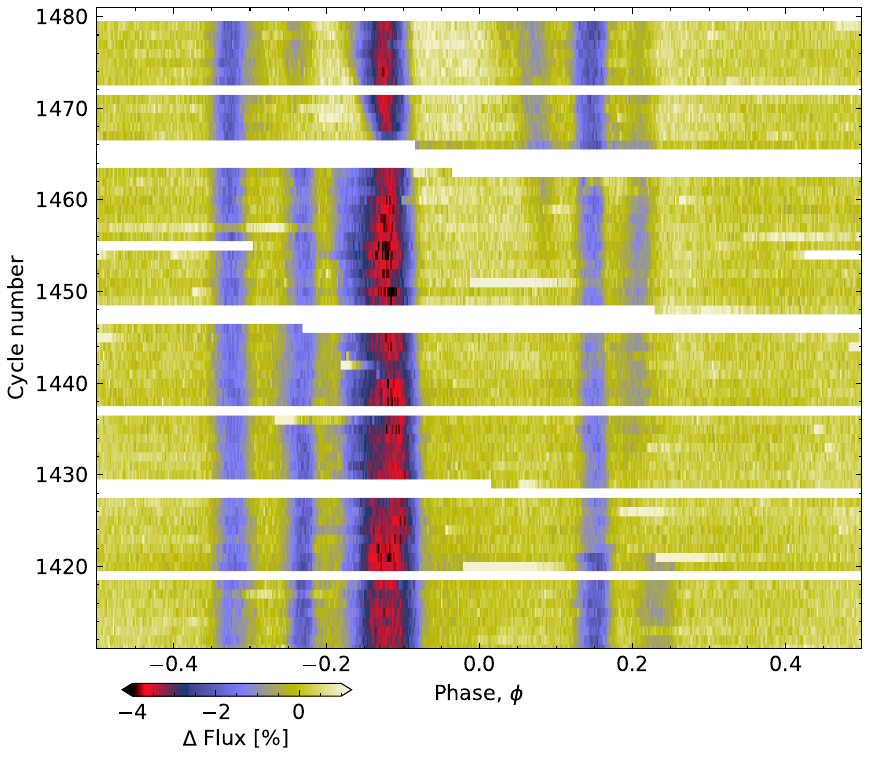}
			\includegraphics[width=0.49\textwidth]{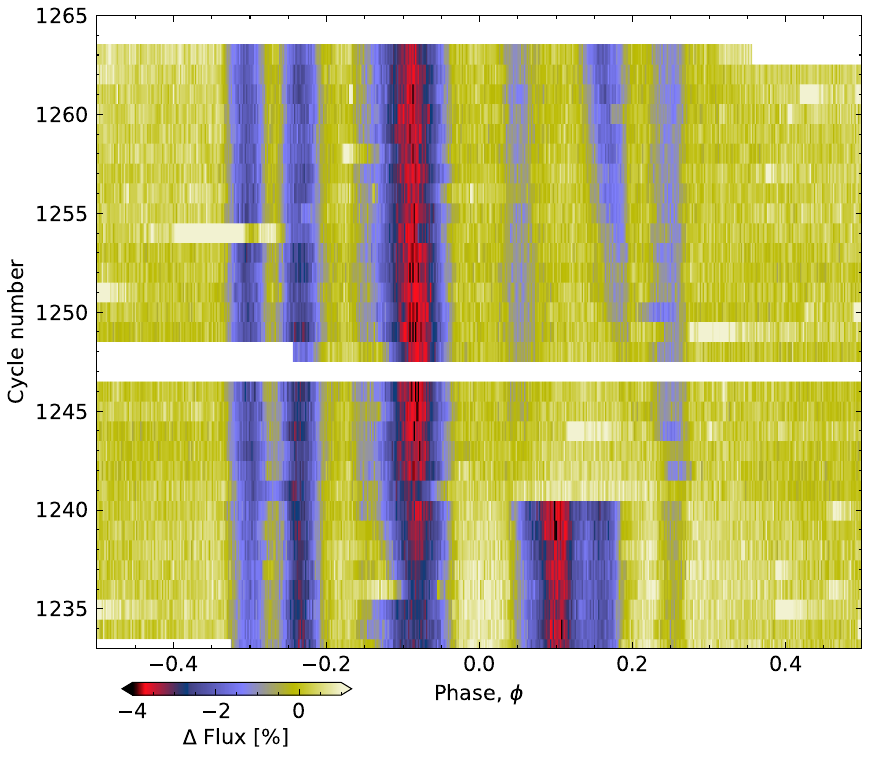}
		}
	\end{center}
	\vspace{-0.4cm}
	\caption{
    {\bf River plots of the LP 12-502 light curve}, showing (clockwise
    from top-left) Sectors 18-19, 25-26, 53, and 58-59.  A
    two-harmonic sinusoid has been subtracted to highlight the sharp
    dips.  A fixed period and phase are adopted for all sectors; the
    dips across all observations are bounded by $\phi \in
    [-0.35,0.35]$.  In Sectors 25-26 (cycles 248-315), periods are
    visible at the fundamental period of 18.5611\,hr, as well as at
    faster ($\phi$$\approx$0-0.07) and slower
    ($\phi$$\approx$0.25-0.27) relative periods based on the presence
    of blue dips with distinct slopes.  Multiple simultaneous periods
    are also visible in Sector 53 (cycles 1234-1263) and Sectors 58-59
    (cycles 1411-1479).  White chunks denote missing data.  The state
    changes noted with red markers in Figure~\ref{fig:lplc} occur in
    cycles 261, 309, and 1241.
	}
	\label{fig:lpriver0}
\end{figure*}

\subsection{Evolution over consecutive sectors, \& LP 12-502}

A few of our complex periodic variables were near the TESS continuous
viewing zones (Figure~\ref{fig:catalogscatter}, top right).  Out of
this already small sample, LP 12-502 (TIC~402980664; $d$=21\,pc,
$J$=9.4, $T$=11.1) stood out due to the quality and content of its
data.  We discuss another interesting source, TIC~300651846, in
Appendix~\ref{app:tic3006}.  In this section, we describe the LP
12-502 observations and the possible implications.

\subsubsection{LP 12-502 observations}
\label{subsec:lpobservations}

Whenever LP 12-502 was located within a TESS sector, it was observed
at 2-minute cadence.  Figure~\ref{fig:lplc} shows all the available
data, from Sectors 18, 19, 25, 26, 53, 58, and 59. Vertical offsets
were applied to separate the data from different spacecraft orbit
numbers; there are always two orbits per sector.  We binned the light
curve to 15-minute intervals to facilitate visual inspection.  Points
more than 2.5$\sigma$ above the median are drawn in gray, to prevent
outliers from seizing attention.  Data gaps are not connected by lines
(a common source of confusion in light curve visualization).
Figure~\ref{fig:lp} shows the same data after phase-folding each TESS
spacecraft orbit, assuming $P$=18.5611\,hr and a fixed reference epoch
of BTJD=1791.5367.  Finally, Figure~\ref{fig:lpriver0} shows ``river
plots'' of the same data, split into similar intervals: the Sector
18-19 data, 25-26 data, 53 data, and 58-59 data.  The river plots are
subject to one additional processing step: we fitted and subtracted a
maximum-likelihood two-harmonic sinusoid independently from the Sector
18-19 data, 25-26 data, and 53, 58, and 59 data in order to accentuate
changes in the dip timing and structure.

The average period, determined by measuring the PDM peak period over
each sector independently, was $\langle P \rangle = 18.5560$\,hr.  The
range between the maximum and minimum sector-specific periods was
measured to be about one minute.   However, a period shift of
$\pm$1\,minute leads to large phase drifts over the entire timespan of
observations.  One minute is $\approx$1/1000$^{\rm th}$ of a period,
and we have observed 1500 cycles.  By folding with a fine grid of
trial periods, we found that the choice $P=18.5611 \pm 0.0001$\,hr causes
more of the features in the LP 12-502 light curve to maintain constant
phases over the entire dataset.

We now attempt to describe the complex morphology of the light curve
and its evolution.  For the first 64 cycles, the star shows four
obvious local minima.  We dub these dips $\{ 1, 2, 3, 4 \}$ at phases
$\{ -0.28, -0.08, 0, 0.25 \}$, respectively.  Dips 2 and 3 are part of
the same ``global'' minimum, which otherwise resembles a long eclipse.
Over cycles 0-64, the depth of dips 1 and 3 remain roughly fixed.  Dip
4 decreases in depth by about 2\%, and dip 2 increases in depth by
about the same amount (see Figure~\ref{fig:lp}).  A subtle fifth dip
may also be present at phase +0.08, at the end of the global minimum
that includes dips 2 and 3.

There is then a 6-month (184-cycle) gap to Cycles 248-315, which show
two highly structured dip complexes, plus a small leading dip.  The
leading dip has the same phase (relative to minimum light) as in
cycles 0-64, and therefore seems likely to be due to the same
structure.  Along a similar line of logic, it seems plausible that the
first ``dip complex'' during cycles 248-264 represents an evolution
and reduction in amplitude of dips 2 and 3 that were seen during
cycles 0-64.  During cycles 266-310, an additional local minimum
develops between the two complexes; this feature is best visualized on
the river plots (Figure~\ref{fig:lpriver0}), where it is seen to have
a shorter period than the other dips (as described below).

The second dip complex during cycles 248-315 shows the most
substructure.  During e.g.~cycles 283-298, this single complex shows
six local minima.  The first and deepest dip is sharp: it shows a flux
excursion of 3.5\% over about 22 minutes (0.02\,$P$), which is the
steepest slope exhibited anywhere in the LP~12-502 dataset.  After the
sharp dip, there is a roughly exponential return to the baseline flux
spanning about a quarter of a period, punctuated by coherent local
minima and maxima that the river plot (Figure~\ref{fig:lpriver0})
reveals to have slightly longer periods than the sharp dip.  The sharp
leading dip remains roughly constant in amplitude until a sudden
``state change'' at BTJD 2030.7 (cycle 309) that occurred at the same
time as a flare, and left the trailing dips seemingly unaffected.
This apparent state change, and two others, are marked with red lines
in Figure~\ref{fig:lplc}.

The behavior during Sectors 53--58 (cycles 1233-1481) is comparatively
tame; the light curve shows only four to six dips per cycle.  Some
dips remain stable in depth and duration over this five-month
interval.  Other dips grow, like the one at $\phi = +0.06$ between
cycles 1458 and 1481.  Other dips, such as the one at $\phi = +0.12$
in cycles 1233-1264, disappear entirely.  The most dramatic state
change occurs during cycle 1241, when a large dip switches from a
depth of 3\% and a duration of 3 hours to a depth of 0.3\% and a
duration of 1 hour.

\subsubsection{Lessons from LP 12-502}
\label{subsec:lplessons}

{\sc State-changes reveal dip independence}---The state-changes seen
in cycles 261, 309, and 1241 confirm that dips can disappear in less
than one cycle. While such behavior was also noted by
\cite{2017AJ....153..152S}, the data presented here show further that
the dips can be {\it independent} and {\it additive}.  For example,
throughout cycles 1233-1264, there are three sharp dips between phases
of 0 and 0.3 with different amplitudes but similar slopes.  During the
transition, the leading dip nearly disappeared while the other two
dips hardly changed; compare the centermost two panels of
Figure~\ref{fig:lp}.  Evidently, the material or process responsible
for one dip can vary independently of the materials or processes
responsible for other dips.  The state changes during cycles 261 and
309 support the same conclusion, while also hinting that the {\it
leading} dip of a complex is most prone to disappearing, leaving the
trailing dips unchanged in its wake.

{\sc Slow growth; rapid death}---LP 12-502 shows at least three
instances in which dips switch off over less than one cycle; we did
not see any such instances of dips switching on.  Dip growth seems to
happen more slowly.  For instance, the dip at phase 0-0.1 between
cycles 258-290 begins to become detectable during cycle 258, and
growths in depth by about 2\% over the next eight cycles.  The
evolution of this particular dip is most clear in the river plots.
The evolution of the dip group at phases 0.1-0.3 during cycles
1410-1481 is another example of this slow mode of dip growth.

{\sc Dip durations}---The shortest dip duration for any of the
individual LP 12-502 dips seems to be $\approx$0.06\,$P$ $\approx$
1.08\,hr.  This is very similar to the characteristic timescale of a
transiting small body at the corotation radius,
\begin{equation}
T_{\rm dur} \equiv R_\star P_{\rm rot} / (\pi a) = 1.02\pm 0.10\,{\rm hr},
\end{equation}
where we have inserted the stellar radius and mass derived in
Section~\ref{subsec:starprops}.  Thus, the shortest-duration dips are
likely produced by transits of bodies or distributions of material
that are smaller than the star.  The corotation radius corresponds to
$a/R_\star \approx 5.8$, i.e., the transit of a body at the corotation
radius has a duration about six times shorter than a feature on the
stellar photosphere that is carried across the visible hemisphere by
rotation.  On the other hand, some dip durations are sufficiently long
that an explanation involving transits would require structures that
are larger than the star along the direction of orbital motion.

{\sc Dip periods}---Most of the LP 12-502 dips repeat with a period of
$P=18.5611 \pm 0.0001$\,hr.  However the river plots
(Figure~\ref{fig:lpriver0}) reveal that a few dips have detectably
distinct periods.  For instance, in sectors 25-26, the dip that
develops around cycle 262 has a period shorter than the mean period by
$\approx$0.1\%, and some of the trailing local minima in the main dip
complex have periods slower than the mean period by $\approx$0.04\%.
In addition to the fundamental period, we were able to identify at
least four distinct periods shown by specific dips over the full
Sectors 18-59 dataset: 18.5683, 18.5672, 18.5473, and 18.5145\,hr,
with a measurement uncertainty of $\approx$0.0002\,hr. Possibly, the
different periods belong to clumps of dust or prominences of gas at
slightly different orbital distances surrounding the corotation
radius.

\section{Discussion}
\label{sec:discussion}

\subsection{Typical and extreme CPVs}
\label{subsec:extreme}

Referring back to Figure~\ref{fig:catalogscatter},  typical CPV masses
span 0.1-0.4\,$M_\odot$, typical ages span 2-150\,Myr, and relative to
the Pleiades, the CPVs are among the more rapidly rotating half of M
dwarfs.  The CPV mass and age range includes both fully convective
stars and stars with a combination of radiative cores and convective
envelopes; the dividing line for these ages is at around $M_\star =
0.25\,M_\odot$ \citep{2018A&A...619A.177B}.  We found no obvious
differences in light curve morphology for CPVs above and below this
fully-convective pre-main-sequence boundary.  

The closest CPV in our catalog is DG~CVn (TIC~368129164), a member of
AB Dor at $d$$=$18\,pc.  The three brightest CPVs are DG~CVn
($T$=9.3), TIC~405754448 ($T$=9.6), and TIC~167664935 ($T$=10.3).  The
shortest period, 3.64\,hr, belongs to TIC~201789285.  The longest
period, 37.9\,hr, belongs to TIC~405910546.  Based on the Gaia$_{\rm
DR3}$ RV scatter, the latter source may turn out to be an eclipsing
binary; if so, the longest-period CPV in our catalog would be
TIC~193831684 (31.0\,hr).  By definition, we required the periods to
be below 48\,hr.

The lowest mass ($\approx$$0.12$\,$M_\odot$) belongs to TIC~267953787.
The catalog contains a few other stars with similar mass.  We cannot
rule out the possibility that CPVs exist with even lower masses, given
the small number of such low-mass stars in our target sample.  Perhaps
even brown dwarfs can be CPVs, although it might be difficult to
distinguish the type of variability we associate with CPVs from the
usual variability of brown dwarfs caused by clouds and latitudinal
bands \citep[e.g.][]{2021ApJ...906...64A,2022ApJ...924...68V}.

\subsection{Is binarity important for CPVs?}
\label{subsec:discbinary}

For CPVs, binarity seems to provide either nuisances or curiosities.
The nuisances include astrophysical false positives with two beating
rapidly rotating stars, as well as uncertainty about which star
produces the CPV signal in binary systems.  Our two candidate K dwarf
CPVs suffer from this latter concern (see
Section~\ref{subsec:massive}).

Curiosities include the four binary systems that are now known to each
host two separate CPVs \added{(see Section~\ref{subsec:multis})}.  CPVs are sufficiently rare that such systems
may have physical import.  Recent work has shown that the orbits of
binaries closer than $\lesssim$700 AU tend to be aligned with their
planetary systems \citep[e.g.][]{2022AJ....163..207C}.  If we assume
that observing CPV variability requires high line-of-sight
inclinations, and that the inclinations in binaries are correlated,
then we would expect the detection of one CPV in a binary system to
raise the probability that the other star is a CPV.  The limitations
of the current catalog prevent further exploration of this issue, but
it might be interesting for future study.

\deleted{A separate possible relation between binarity and CPVs is that the
presence of a binary companion could cause early disk dispersal,
freeing the star to contract.  The CPV phenomenon seems to require
rapid rotation; thus if binary systems tend to produce more rapidly
rotating stars, we would expect a sample of CPVs to have a larger
binary fraction than the field.  However, we see minimal evidence for
such an effect (Section~\ref{subsec:resultsbinarity}).  The
multiplicity rate of M dwarfs near the Sun is 26.8$\pm$1.4\%
\citep{2019AJ....157..216W}.  The $\approx$32\% multiplicity fraction
in our CPV sample (from RUWE$_{\rm DR3}$; see
Section~\ref{subsec:resultsbinarity}) seems approximately consistent
with this field fraction.}

\subsection{Transience of CPV dips}

While CPV periods appear to remain fixed over thousands of cycles, the
light curve shapes evolve over typical timescales of 10 to 1{,}000
cycles (e.g.\ Figures~\ref{fig:evoln} and~\ref{fig:lpriver0}).
Although we refer to them as ``periodic'', the CPVs are therefore
actually quasiperiodic, with coherence timescales of $\approx$100
cycles.  This marks a qualitative departure from the ``persistent''
{\it vs.} ``transient'' flux dip distinction previously described by
\citet{2017AJ....153..152S}, which was based on $\lesssim$100\,cycle
K2 baselines.  The observation that CPVs have a population-averaged
on-off duty cycle of $\approx$75\% (Figure~\ref{fig:evoln}) is also
new.  Appendix~\ref{app:tic3006} for instance shows $\approx$1{,}000
cycles of a source, TIC~300651846, with between zero and five sharp
local minima per cycle.  During the ``zero'' epochs (cycles
$\approx$503-542), the source would likely be labeled an ordinary
rotating star.

\subsection{Special phases of CPV dips}

An independent peculiarity of CPV evolution is that the dips do not
explore all phase angles with equal weight.  LP 12-502, and other
CPVs, exhibit preferred phases lasting for at least two years.  For LP
12-502, all of the dips happen over phases corresponding to only two
thirds of the period (Figures~\ref{fig:lp} and~\ref{fig:lpriver0}).
The remaining third seems to be ``out of limits'' for dips over the
timespan of observations.  This could be evidence that the stellar
magnetic field is not azimuthally symmetric.  Alternatively, the
source of the material (e.g. a planetesimal swarm) might be
distributed over an arc rather than occurring randomly around the
entire orbit.

\subsection{Dip asymmetries?}

The asymmetry of a dip around the time of minimum light might be
caused by the variation in optical depth of the occulting material as
a function of orbital phase angle.  Sharp leading edges with trailing
exponential egresses, for instance, have been previously seen for
transiting exocomets and disintegrating rocky bodies
\citep[e.g.][]{2012ApJ...752....1R,2012A&A...545L...5B,2015Natur.526..546V,2019A&A...625L..13Z}.

Examining Figure~\ref{fig:cpvs}, it is clear that CPV dips can be
asymmetric but it is not obvious whether there is a preference for
sharper ingresses or sharper egresses.  In some cases (e.g.
TIC~425933644), the flux variations do not resemble isolated dips,
making the meaning of ``ingress'' and ``egress'' unclear.  In other
cases, such as Sector~36 of TIC~89463560, there is a sharp drop with
an exponential return to the baseline flux, resembling the signatures
of exocomets \citep[e.g.][]{2018MNRAS.474.1453R,2019A&A...625L..13Z},
and the outflowing exospheres of some transiting planets
\citep[e.g.][]{2019ApJ...873...89M,2022ApJ...926..226M}.

\subsection{What causes the CPV phenomenon: dust vs.\ gas}

Both the dust clump and the gas prominence scenarios
(Figure~\ref{fig:f1} and Section~\ref{sec:intro}) invoke clumps of
material at the corotation radius; one property that distinguishes the
two ideas is the composition of the material.

\subsubsection{What is a prominence?}
The prominence idea is based on a loose analogy with quiescent
prominences/filaments in the solar corona that last as long as a few
weeks \citep[see][]{2015ASSL..415.....V}.  In the context of the Sun,
a prominence is a clump of cold, partially ionized hydrogen viewed in
emission against the dark backdrop of space.  A filament is the same
clump of plasma, but viewed in absorption against the solar disk.  In
an extrasolar context, spectroscopic detections of transient Balmer-
and resonance-line absorption seen for stars such as AB~Dor and
Speedy~Mic
\citep[e.g.][]{1989MNRAS.238..657C,1993MNRAS.262..369J,2006MNRAS.365..530D,2016MNRAS.463..965L}
have been interpreted as prominences that scatter a star's
chromospheric emission \citep[see][]{1989MNRAS.238..657C}.  The
short-term mechanical stability of such gas configurations is
theoretically plausible for rapid rotators
\citep{2000MNRAS.316..647F,2022MNRAS.514.5465W}.   To our best
knowledge, this class of spectroscopic observation also has no viable
alternative explanations.

We performed a simple visual examination of the TESS light curves for
five prominence-hosting systems studied by
\citet{2019MNRAS.482.2853J}--AB~Dor, Speedy~Mic, LQ Lup, HK Aqr, and
V374 Peg--and detected no CPV behavior.  While individual prominences
may only last one to tens of rotation cycles, the prominence system
itself is thought to always be ``on'', due to the repeatable
detectability of spectroscopic transients \citep[e.g.][and references
therein]{1990MNRAS.247..415C}.  Assuming that spectroscopically
observable prominence systems indeed do not turn off, this would imply
that they are not always accompanied by photometric CPV-like dips: a
link between the spectroscopic prominences that may exist around
rapidly rotating low-mass stars and the CPV phenomenon has yet to be
made.

\subsubsection{What is the microphysical source of opacity?}

CPVs show broadband flux variations that can be 1-2$\times$ deeper in
the blue than in the red
\citep{2017PASJ...69L...2O,2020AJ....160...86B,2022AJ....163..144G,2023MNRAS.518.2921K}.
Dust can naturally explain this chromaticity, since it has a larger
absorption cross-section in the blue than the red
\citep[e.g.][]{1989ApJ...345..245C}.  Gas might also explain the
observed chromaticities \citep{1992oasp.book.....G}.  While
bound-bound absorption can be excluded, since it provides opacity only
at narrow resonant lines, the hydrogen opacity due to bound-free
absorption is ``jagged'' \citep[see][Figure 8.5 and
Eq.~8.8]{1992oasp.book.....G}, such that at temperatures of
$\approx$3{,}000\,K to $\approx$10{,}000\,K the opacity can be larger
at blue wavelengths than at red wavelengths.  Bound-free absorption of
${\rm H}^{-}$ is often important at such temperatures, but this
opacity source is stronger in the red than the blue, the opposite of
what is required to produce deeper dips in the blue than in the red.
Likewise, Thomson scattering is too gray to be the dominant opacity
source.  From hydrogen alone, bound-free absorption therefore seems
like the most plausible opacity source.  However it remains to be
demonstrated whether a sufficient population of excited states could
be maintained, particularly given the short ($\approx$microsecond)
radiative decay timescales.

An instructive point of comparison is the rapidly rotating magnetic B
star, $\sigma$~Ori~E, which shows dips that are deeper in the blue
than in the red \citep{1977ApJ...216L..31H}.  Photometric and
spectroscopic observations of this star have been understood in terms
of a warped torus of corotating circumstellar material
\citep{1978ApJ...224L...5L,1985Ap&SS.116..285N,2005ApJ...630L..81T}.
The circumstellar material is unlikely to be dust, which would
sublimate quickly at the distance of the torus from the
star.\footnote{\citet{2019ApJ...876..127Z} explored the sublimation
timescales for a canonical CPV with $M_\star=0.2\,M_\odot$,
$R_\star=0.3\,R_\odot$, and $T_{\rm eff}=3200\,{\rm K}$.  They found
that non-shielded, generic silicate dust mixture
\citep{1985ApJS...57..587D} with a single size of 0.1\,$\mu$m reached
the $\approx$1500\,K sublimation temperature at $\approx$3\,$R_\star$.
This suggests that dust sublimation could be an important effect even
for CPVs. }  The opacity source for $\sigma$~Ori~E and its analogs is
instead thought to be bound-free absorption by neutral hydrogen
\citep{1985Ap&SS.116..285N}, although to our best knowledge direct
evidence for this conclusion has yet to be acquired.  Separate and
smaller-amplitude continuum flux brightenings in $\sigma$~Ori~E may
also come from electrons scattering photospheric light toward the
observer when the clouds are not transiting
\citep{2022MNRAS.511.4815B}.

Given these complexities, it seems important for a future theoretical
study to be conducted to determine to what degree the observed
chromaticities in CPVs match, or do not match, expectations from
radiative transfer.  This issue has a key ability to resolve the
question of whether the CPVs are explained by dust or by gas, which
has bearing on whether the material producing the dips is coming from
the star, or whether it is a byproduct of the protoplanetary disk.

\subsubsection{The lifetime constraint}

The observed lifetime of the CPV phenomenon could provide another way
to discern between the gas and dust clump scenarios.  Based on the
available statistics from \deleted{e.g. }\citet{2022AJ....164...80R} and
references therein, it seems plausible that CPV occurrence decreases
with stellar age from $\approx$3\% at 10\,Myr (Sco-Cen), to
$\approx$1\% at 100\,Myr (Pleiades), down to 0\% by the
$\approx$700\,Myr age of Praesepe.  This is odd in the context of the
prominence scenario, because pre-main-sequence M dwarfs spin up over
the first 100\,Myr; prominences might therefore be expected to be {\it
more} common at 100\,Myr than at 10\,Myr, under the assumption that
the production of prominences depends only on the stellar rotation
rate.  The dust clump scenario would hold a natural explanation: the
lower occurrence of CPVs around older stars would simply reflect a
finite supply of dust.  One \replaced{possible}{potential} complication however is that the
magnetic field topology of rapidly rotating M dwarfs may depend on
factors other than the rotation rate (e.g.\ the age), which might
alter the production of prominences.

\added{\subsection{Why are the dip and spot periods nearly equal?}}

\added{The CPV dips are usually superposed on nearly sinusoidal modulation.
The sinusoidal modulation
is probably induced by brightness inhomogeneities on the stellar surface.  If
the dips are explained by material orbiting the star, then the proximity
of the spot and dip periods is surprising.
For instance, the dust clumps modeled by \citet{2023MNRAS.518.4734S} tend to
accumulate near -- but not exactly at -- corotation.}

\added{To explore the proximity of the dip and spot periods, we
generated synthetic light curves that superposed a sinusoid and a single
eclipse.  We imposed periods, amplitudes, sampling, and noise properties
similar to typical CPVs in Figure~\ref{fig:cpvs}.  We then tuned
the period of the eclipse signal to differ slightly from the
sinusoidal period, and processed the resulting synthetic signals
through the same period-finding routine to which we subjected the real data.
Provided the dip and sinusoid periods agreed to within $\approx$0.1\%, we found that
phase-folding on the dominant period (that of the larger-amplitude starspots)
yielded dip signals analogous to those in Figure~\ref{fig:cpvs}.  However,
increasing the period difference beyond $\gtrsim$0.3\%, the dip signal quickly becomes
``smeared'' out and unidentifiable when phase-folding.  This
exercise highlights that our search was sensitive only to stars with dip
and spot periods that agreed to within a few parts per thousand.  
Some CPVs may exceed this threshold; we encourage
future work aimed at determining whether such systems exist.}

\subsection{Planets or planetesimals near corotation?}

Close-in planets are common around M dwarfs; studies from Kepler have
shown that early M dwarfs have $\approx$0.1 planets per star with
sizes between 1-4\,$R_\oplus$ and orbital periods within 3 days
\citep{2015ApJ...807...45D}.  The frequency of planets per star
increases to $\approx$0.7 when considering periods as long as
10\,days.  Extrapolating to all small (0.1-4\,$R_\oplus$) planets
within 10 days, it is reasonable to expect nearly all M dwarfs to have
at least one planet.

In the context of disk-driven planet migration, the stopping location
for the innermost planet is set by the protoplanetary disk's
truncation radius \citep[e.g.][and references
therein]{2018haex.bookE.142I}.  The truncation radius is often
calculated by equating the magnetic pressure from the stellar
magnetosphere with the ram pressure of the inflowing gas.  As it
happens, the truncation radius is close to the corotation radius for
low accretion rates
\citep[e.g.][]{2015SSRv..191..339R,2022MNRAS.510.5246L}.  These
considerations invite us to imagine one or more planets migrating
inward due to gas drag, and arriving at $\approx$5-10 stellar radii
before the disk is depleted.

With this picture in mind, it is tempting to attribute features of the
CPV light curves to transits of material ejected by planets or
planetesimals.  Young rocky bodies are expected to be hot, and they
might expel either gas or dust.  The Jupiter-Io system
\citep[e.g.][]{2004jpsm.book..537S} is analogous, in that a small
rocky body feeds the construction of a plasma torus.  We emphasize
that although this type of configuration seems a priori plausible, no
direct evidence currently supports it.

The main logical function of the planetesimals would be to serve as a
source for the occulting gas or dust; they would not necessarily need
to explain the observed phases of the observed dips.  The azimuthal
angle of the eventual entrainment could be entirely dictated by the
stellar magnetic field.  In this scenario, the obscuring material
would inspiral from one or more rocky bodies well beyond the
corotation radius.  The planetesimals themselves would not necessarily
need to transit.  However if they did, they would need to be $\lesssim
1$\,$R_\oplus$ based on their non-detections in the TESS data.
Possibly analogous systems include K2-22 \citep{2015ApJ...812..112S}
and KOI-2700 \citep{2014ApJ...784...40R}, though the obscuring
material in the CPVs would need to be observed much further from the
emitting planet than for those two examples.  

A more restrictive variant of the planetesimal scenario would be to
posit that the obscuring material remains close to the launching body,
similar to comets, or to the aforementioned K2-22 and KOI-2700
systems.  If so, then the planetesimals would need to be at the
corotation radius.  One prediction would therefore be that certain
orbital phases would produce recurrent dips when observed over
sufficiently long baselines, because the launching planetesimal would
be massive enough to remain in orbit, while stochastically ejecting
material.  For most CPVs (Figure~\ref{fig:evoln}), the data seem to be
in tension with this expectation because the relative spacing between
dips is almost never conserved.  With that said, certain sources do
seem to exhibit ``special phases'', including LP 12-502
(TIC~402980664), DG~CVn (TIC~368129164), TIC~193831684, and
TIC~146539195.  One possible explanation for this might be if
obscuring material is remaining close to its launching body, or
bodies.  An alternative explanation could be that the stellar magnetic
field configurations responsible for confining said material are
stable over the existing two-year baseline.

\subsection{Mass flux estimate}
\label{subsec:massflux}

Assuming for the moment that the obscuring material is dust, we can
estimate the mass of a transiting clump. First, we convert the transit
depth into an effective cloud radius, $R_{\rm cloud}$.  For most CPVs
in Figure~\ref{fig:cpvs}, this yields $\approx$2-20\,$R_\oplus$.  A
minimum constraint on the number density of dust particles is obtained
by requiring the cloud to be optically thick.  For cases like LP
12-502, this is reasonable because the transit duration of the
shortest dips implies $R_{\rm cloud}\ll R_\star$.  Carrying out the
relevant calculation assuming the dust grains are 1\,$\mu$m in size,
\citet{2023MNRAS.518.4734S} reported minimum cloud masses of order
$10^{12}$\,kg (their Eq.~23), which scale linearly with both the
optical depth and dust grain radius.  This is comparable to a small
asteroid; the asteroid belt itself has a mass of order
$\approx$10$^{21}$\,kg \citep{2019Icar..319..812P}.  A similar
calculation that assumed occulting clumps of hydrogen, rather than
dust, derived gas prominence masses of at least $10^{14}$\,kg
\citep{1990MNRAS.247..415C}, about 100$\times$ larger than the lower
limit on the dust mass.

If the disappearance of a dip represents the permanent loss of the
obscuring material -- for example, if it is the result of a dust clump
being accreted or ejected -- then we can also estimate the rate at
which mass is flowing through the structures that lead to dips.  For
instance, LP~12-502 showed three ``state-switch'' events over the six
months of available TESS observations, during cycles 261, 309, and
1241 (Figure~\ref{fig:lpriver0}).  The other source for which we
performed a comparable analysis, TIC~300651846
(Appendix~\ref{app:tic3006}), showed two state-switches over 11
months.  In all such cases, at least one dip turned off.  For purposes
of estimation, we will take LP 12-502 as our prototype.  Assuming the
occulting material is dust, the corresponding $\dot{M} \equiv M\cdot
{\rm d}N/{\rm d}t$ time-averaged over six months is
$\approx$$1\times10^{-12} M_\oplus\,{\rm yr}^{-1}$.  Considered
cumulatively over the $\approx$$10^8$ years for which the CPV
phenomenon is observed, this yields a cumulative moved dust mass of
$10^{-4}\,M_\oplus$, of order the Solar System's asteroid belt.  If
the occulting material is gas, the lower mass bounds would be of order
100 times larger.  For cases in which we observe the {\it growth} of
dips, such as the Sector~29 data for TIC~224283342, or Sector~5 of
TIC~294328885, the dip depths typically increase by of order a few
percent over ten to twenty days.  This growth rate yields a mass flux
one order of magnitude larger than the earlier estimate.

\subsection{From dippers to debris disks}
\label{subsec:discdippers}

About one in three young stars with infrared-detected inner dusty
disks show quasiperiodic or stochastic dimming over timescales of
roughly one day
\citep[e.g.][]{2010A&A...519A..88A,2010ApJS..191..389C}.  The dimming
amplitudes can reach a few tenths of the stellar brightness, and dips
with identical depths and phases rarely recur.  These ``dipper'' stars
are probably explained by occulting circumstellar dust in the inner
disk
\citep[e.g.][]{2014AJ....147...82C,2016ApJ...816...69A,2021ApJ...908...16R,2022ApJS..263...14C}.
While the phenomenon can persist beyond $\approx$10\,Myr
\citep{2019MNRAS.488.4465G,2022MNRAS.514.1386G}, in all such cases it
seems to be associated with the presence of infrared excesses.
Phenomenologically, dippers are different from CPVs in that their dips
are usually deeper, less periodic, and more variable in depth over
timescales of only one or a few cycles.  Dipper stars also tend to be
younger, since they tend to be classical T Tauri stars with infrared
excesses.

In identifying the two candidate CPVs with outlying SEDs
(TICs~193136669 and TIC~57830249; Section~\ref{subsec:irexcess}), we
were prompted to reconsider our light curve-based labeling, and
ultimately concluded that these sources are dippers.  This episode
suggests that there could be overlap between CPVs and dippers.  Taking
TIC~57830249 as one example, the Sector~36 TESS data are suggestive of
a CPV, with relatively periodic, sharp dips with depths of a few
percent.  The Sector~10 data are completely different, varying in
apparent flux by a factor of two, with no discernible periodicity at
all.  Perhaps this source becomes a ``dipper'' when an inflow of dust
reaches the inner disk wall, and is otherwise a ``CPV'' when the inner
disk is starved of dust.

Although TIC~57830249 is an intriguing outlier, the general picture is
that stars without infrared excesses have more stable optical light
curves than those with infrared excesses.  While some dippers may
evolve into CPVs after the disk is mostly gone, this would be
generically expected based on population statistics: young objects
become old.  There may be no other causal connection between the two
evolutionary stages.  With that said, a common mystery between the
CPVs and dippers is how exactly the {\it narrowness} of their flux
dimmings is produced.  A similar mechanism may operate for both types
of object, tied perhaps to a shared magnetic topology, or perhaps to a
preference for dust to inspiral to the star in clumped structures.

\begin{figure}[!th]
	\begin{center}
		\centering
		\subfloat{
			\includegraphics[width=0.48\textwidth]{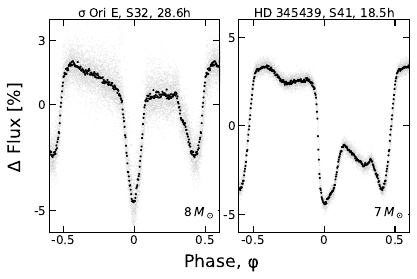}
		}
			
		\vspace{-0.34cm}
		\subfloat{
			\includegraphics[width=0.48\textwidth]{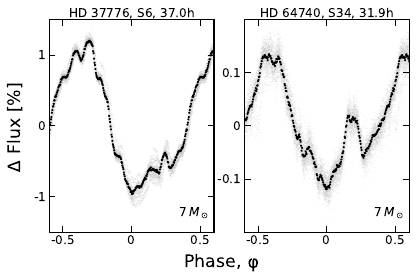}
		}

		\vspace{-0.34cm}
		\subfloat{
			\includegraphics[width=0.48\textwidth]{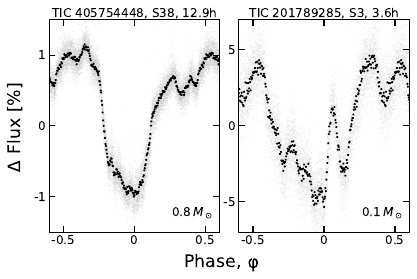}
		}

		\vspace{-0.1cm}
		\caption{
      {\bf The magnetic B star connection.}
      $\sigma$~Ori~E and HD~345439 (top row) are magnetic B stars with
      predominantly dipolar magnetic fields known to host
      circumstellar plasma tori.  HD~37776 and HD~64740 (middle row)
      are analogous magnetic B stars with field topologies potentially
      dominated by high order multipoles.  The bottom row compares the
      latter systems against the ``best-matching'' CPV light curves,
      selected by eye from Figure~\ref{fig:cpvs}.  CPVs have light
      curves that are visually similar to the topologically complex
      magnetic B stars.  Stellar masses rounded to one significant
      figure are given in the lower right of each panel; the star,
      TESS sector, and period are listed in each subtitle.
		}
		\label{fig:bstar}
	\end{center}
\end{figure}

\subsection{Strengthening the magnetic B star connection}
\label{subsec:bstardisc}

\citet{2017AJ....153..152S} previously noted a possible connection
between the CPVs and rapidly rotating magnetic B stars such as
$\sigma$~Ori~E, which can have circumstellar gas clouds trapped in
corotation \citep{2005ApJ...630L..81T}.  The $\sigma$~Ori class is
distinct from Be-star decretion disks, which are found to
systematically not host detectable magnetic fields
\citep{2013A&ARv..21...69R,2016ASPC..506..207W}.

An argument against the connection between CPVs and the $\sigma$~Ori~E
analogs is that the light curve of $\sigma$~Ori~E is simpler than
those in Figure~\ref{fig:cpvs}, with only two broad local minima, and
one ``hump'' (Figure~\ref{fig:bstar}; see also
\citealt{2022ApJ...924L..10J}).  Within the model proposed by
\citeauthor{2005ApJ...630L..81T}, the simplicity of the light curve is
the result of a simple dipolar magnetic field, which is typical of
magnetic B stars \citep{2007A&A...475.1053A,2009ARA&A..47..333D}.  The
magnetic axis needs to be tilted relative to the stellar spin axis in
order to match the qualitative behavior of both the broadband light
curves, and the line-profile variations seen in hydrogen, helium, and
carbon \citep{2012MNRAS.419..959O}.

Two interesting and possibly telling exceptions to the rule that
magnetic B stars have simple light curves are HD~37776 and HD~64740.
HD~37776 is known from spectropolarimetry to have an extreme field
geometry dominated by high order multipoles
\citep{2011ApJ...726...24K}.  The field geometry of HD~64740, while
potentially less extreme, also shows evidence for a non-dipolar
contribution \citep{2018MNRAS.475.5144S}.  Recent TESS light curves of
these two B stars appear surprisingly similar to the CPV light curves
\citep{2020pase.conf...46M}.  The middle row of Figure~\ref{fig:bstar}
shows the phased TESS light curves for these two stars, with by-eye
best-matching CPVs shown underneath for comparison.  The number of
dips per cycle, the shapes of the dips, and the dip depths relative to
the sinusoidal envelope are all similar.  This connection suggests
that the highly structured light curves of both the M dwarfs and the B
stars are associated with (and perhaps caused by) strong non-dipolar
magnetic fields.  Non-dipolar fields for M dwarfs are plausible, given
that Zeeman Doppler Imaging has revealed non-axisymmetric magnetic
field patterns for the few M dwarfs for which this technique is
technically feasible \citep[see][and references
therein]{2021A&ARv..29....1K}.

The physical similarity between the B stars and the M dwarfs could
have its origin in the existence of a ``centrifugal magnetosphere''
\citep[see][]{2013MNRAS.429..398P}.  In other words, both classes of
objects might satisfy the condition $R_{\rm m} > R_{\rm c}$, for
$R_{\rm m}$ the magnetosphere radius (sometimes called the Alfv\'en
radius).  Provided that charged particles are confined to move along
magnetic field lines, material can then build up at the corotation
radius \citep[e.g.][Sec.\ 4, and references
therein]{2015SSRv..191..339R}.  In the converse ``dynamical'' case,
when $R_{\rm m} < R_{\rm c}$, material interior to the magnetospheric
radius returns to the stellar surface over the free-fall timescale.  A
simple estimate assuming a dipole field with $B_0\approx1\,{\rm kG}$
at the star's surface, a local plasma number density
$n\approx10^9\,{\rm cm^{-3}}$, and a plasma temperature $10^6\,{\rm
K}$ gives magnetospheric radii of order a few times the corotation
radii, $R_{\rm c}$.  This suggests that the existence of a centrifugal
magnetosphere is plausible for young, rapidly rotating M dwarfs.

%

\section{Conclusions}
\label{sec:conclusion}

In this work, we searched 2-minute cadence TESS data collected from
2018 July to 2022 September for complex periodic variables (CPVs).
The target stars were \nstarssearched\ late-K and early-to-mid M
dwarfs within 150\,pc and with TESS magnitudes $T$<16.  The selection
function included $>$$80\%$ of such stars within 30\,pc, and $<$$10\%$
of such stars at distances exceeding 100\,pc
(Figure~\ref{fig:completeness}).

We found \ngoods\ objects that showed complex quasiperiodic behavior
over at least one TESS sector.  These \ngoods\ bona fide CPVs are
listed in Table~\ref{tab:thetable}.  This table also includes
\nmaybes\ ambiguous CPVs, whose designation is less certain, and
\ndebunked\ impostors.  We inferred ages for all but two of the
\nallcands\ objects based on memberships in young stellar
associations; we also derived temperatures and radii using SED
fitting, and inferred stellar masses by interpolating against stellar
evolutionary models.  We caution that our sample is far from being
volume-limited and is not even magnitude-limited: the TESS 2-minute
stellar sample had a heterogeneous selection function which may have
been biased in favor of young stars over field stars.  Previous work
however has shown that $\approx$1-3\% of M dwarfs younger than
$\approx$100\,Myr show the CPV phenomenon
\citep{2016AJ....152..114R,2022AJ....163..144G,2022AJ....164...80R}.

Analyzing the TESS light curves and stellar properties of our CPVs, we
draw the following conclusions.

\vspace{-2pt}
\begin{enumerate}[leftmargin=*]
	  \setlength\itemsep{-2pt}
    \item The sharpest CPV dips have durations of $\approx$0.05\,$P$
      and depths of $\approx$1-3\% (Figures~\ref{fig:cpvs}
      and~\ref{fig:evoln}).  Explaining dips this sharp requires
      material extrinsic to the stellar surface (see
      Section~\ref{sec:intro}).
    \item The shortest CPV dips, also with durations of
      $\approx$0.05\,$P$, match the expected transit duration for a
      small body at the corotation radius, $T_{\rm dur} \equiv R_\star
      P_{\rm rot} / (\pi a)$ (see Section~\ref{subsec:lplessons}).
      Such dips are therefore likely produced by transits of bodies or
      distributions of optically-thick material that are smaller than
      the star.
    \item Many CPV dips have durations a few times longer than $T_{\rm
      dur}$ (Figure~\ref{fig:cpvs}).  The dips are often superposed on
      a quasi-sinusoidal signal that presumably originates from
      starspots and faculae on the stellar surface.  The only viable
      explanation currently known for sharp dips being superposed on
      the starspot signals is that concentrations (``clumps'') of
      circumstellar material corotate with the star.  Assuming that
      the longer dips have the same physical origin as the shortest
      dips, the corotating clumps must also be capable of having sizes
      comparable to the star.
    \item The mean periods of CPVs remain fixed to within a relative
      precision $\lesssim$0.1\% over the two-year ($\approx$1{,}000
      cycle) baseline of available observations.  The light curve
      shapes always evolve over this timescale
      (Figure~\ref{fig:evoln}).
    \item The dips in CPV light curves can have slightly different
      periods.  LP 12-502, for instance, showed dips with four
      distinct periods within $\pm 0.3\%$ of its fundamental period,
      sometimes simultaneously, and each lasting for up to 50 cycles
      (Figure~\ref{fig:lpriver0}).
    \item The CPV peaks and dips evolve over timescales that are both
      secular ($\approx$100\,cycles) and impulsive ($<$1 cycle).  Dip
      growth seems to happen over durations of at least ten cycles,
      and slow dip decay can also occur.  ``State-switches''
      correspond to dips collapsing instantaneously;  they occur once
      every few months for both LP~12-502 and TIC~300651846.
      State-switches are almost always linked with observed optical
      flares.  Such switches are suggestive of magnetic reconnection
      opening the ``magnetic cage'' that traps the dust.
    \item The detailed morphology changes exhibited by LP~12-502
      during its state-switches (e.g.\ Figure~\ref{fig:lp}, cycles
      1233-1264) imply that the flux dips are additive and
      independent.
    \item The on-off duty cycle for CPVs is $\approx$$75$\%, based
      on the fraction of bona fide CPVs that either turned on or turned
      off during TESS re-observations, two years after the initial
      observation (Figure~\ref{fig:evoln}).
    \item The CPV phenomenon persists for $\gtrsim$150\,Myr, based on
      the existence of multiple CPVs in AB Dor, the Pleiades, and
      Psc-Eri (Section~\ref{subsec:starprops}).  It may even extend to
      200\,Myr, based on the one CPV we found in the Carina Near
      moving group (TIC~294328887; $\approx$200\,Myr).  The lack of
      detected CPVs in the Hyades and Praesepe suggests that the
      lifetime of the phenomenon is limited to the first few hundred
      million years.
    \item Most CPVs are M dwarfs with masses 0.1-0.4\,$M_\odot$.  Two
      sources, TIC~405754448 and TIC~405910546, have masses that
      appear to exceed 0.5\,$M_\odot$.  Both are potentially binaries,
      and this may confuse our ability to accurate identify the source
      of the CPV signal (Section~\ref{subsec:massive}).  We encourage
      additional scrutiny of these objects in future work.
    \item The closest CPVs to the Sun are at distances of 15-20\,pc;
      the brightest have $V$$\approx$12 ($J$$\approx$7.5).  We have
      found most of the close exemplars in this work, since our CPV
      sample was $\gtrsim$80\% complete within 30\,pc.  The lack of
      CPVs in the volume-complete $<$15\,pc sample of
      0.1-0.3\,$M_\odot$ stars analyzed by \citet{2021AJ....161...63W}
      is consistent with this estimate.  Expanding our analysis of the
      TESS data to the full frame images would yield a truly
      volume-limited selection function, and would expand the CPV
      census by about a factor of two within 50\,pc, and by a factor
      of ten within 100\,pc.
    \item Surprising analogs to CPVs exist in two magnetic B stars,
      one of which is known to have an extreme multipolar field
      topology (Section~\ref{subsec:bstardisc}).  Since most magnetic
      B stars have dipolar magnetic fields, this suggests that the CPV
      dips and warps are similarly being sculpted by the stellar
      magnetic fields, and that the magnetic fields themselves are
      potentially also multipolar.
    \item The rate of dip evolution can be used to place a
      model-dependent lower bound on how much material is either being
      accreted or ejected during the state changes
      (Section~\ref{subsec:massflux}).  Order of magnitude estimates
      require at least an asteroid belt's worth of dust
      ($10^{-4}$\,$M_\oplus$) over $10^8$ years, or at least
      $\approx$$10^{-2}$\,$M_\oplus$ if the occulting material is gas.
\end{enumerate}

While many questions remain, two in particular will be important for
clarifying what these objects might teach us in a broader
astrophysical context: {\it 1)} Is the eclipsing material responsible
for the phenomenon gas or dust?  {\it 2)} What sets the characteristic
clumping size for the circumstellar material?

The distinction between gas or dust is important because it could
clarify whether the CPV phenomenon is intrinsic, so that material
comes from the star, or extrinsic, so that it is sourced through some
generic evolutionary phase of debris disks.  This knowledge would in
turn propagate to our understanding of whether the phenomenon is
primarily teaching us about dust production and processing in gas-poor
disks, or whether it is teaching us about the ability of cold gas to
remain stable in hot stellar coronae for long durations.
Observationally, acquisition of medium- or high-resolution time-series
spectra holds a good chance at resolving the gas vs.~dust question.
Given our observed $\approx$75\% on-off duty cycles, such data must be
acquired simultaneously with photometric time-series observations
(e.g. during TESS re-observation) in order for detections and
non-detections to be interpretable.

In both the gas and dust scenarios, CPVs are preferentially viewed
edge-on.  This implies that after correcting for the line-of-sight
inclination, roughly one third of low mass stars \citep[those that
rotate rapidly enough;][]{2022AJ....163..144G} could trap
circumstellar material in the same way.  It also suggests that CPVs
may preferentially show transiting planets at larger distances than
the corotating material, though this conclusion would be dependent on
whether the magnetic and stellar spin axes tend to be aligned.  Given
these points, observational follow-up work should include searching
for outer transiting planets, and measuring equatorial velocities in
order to test whether the stellar inclination angles are indeed
preferentially edge-on.  Any source of empirical information on the
stellar magnetic field, whether from the Zeeman effect
\citep[e.g.][]{2021A&ARv..29....1K} or perhaps radio emission
\citep[e.g.][]{2015Natur.523..568H}, could also help clarify the
strength of the magnetospheres for these objects.

On the theoretical front, building a physical understanding of what
sets the characteristic size scale of the clumping material would help
clarify why the light curves have the bizarre shapes that are
observed.  The relevant puzzles in plasma physics and radiative
transfer could perhaps be connected to our understanding of the
close-in rocky planets that are expected to be present around most of
these stars.

\acknowledgments
LGB is grateful for support from the Heising-Simons 51~Pegasi~b
Fellowship, and for helpful conversations with J.~Spake, A.~Mann,
G.~Laughlin, and B.~Draine.  
We are also grateful for the assistance of S.~Yee, L.~Weiss,
H.~Isaacson, and A.~Howard in acquiring and reducing the HIRES
spectra\added{, and for the reviewer's suggestion to consider
the proximity of the dip and spot periods}.
This paper relied primarily on data collected by the TESS mission; the
specific 2-minute cadence observations can be accessed via
DOI\,\dataset[10.17909/t9-nmc8-f686]{https://doi.org/10.17909/t9-nmc8-f686}.
Funding for the TESS mission is provided by NASA’s Science Mission
Directorate.  
We thank the TESS Architects (G.~Ricker, R.~Vanderspek, D.~Latham,
S.~Seager, and J.~Jenkins) and the many TESS team members for their
efforts to make the mission a continued success. 
LP 12-502 in particular was observed at 2-minute cadence thanks to the
TESS Guest Investigator programs G022252 (PI: J.~Schlieder; Sectors
18, 19, 25, 26) and G04168 (PI: R.~Jayaraman; Sector 53). 
ADU acknowledges support by NASA under award number 80GSFC21M0002, as
well as ROSES award 22-ADAP22-0070.

LGB conceived the project, performed the dip-counting search, light curve classification, cluster membership, SED, variability, and secondary-period analyses, and wrote the manuscript.
RJ and SR performed the Fourier-based analysis and contributed to light curve classification.
LR cross-examined the light curve classification, and contributed an independent SED analysis.
ADU identified the magnetic B star connection.
LAH contributed to project design.
JNW, SR, and LAH significantly improved the clarity of the manuscript.
G\'AB acquired and maintained the servers used to run the dip-finding pipeline.


\software{
  \texttt{astrobase} \citep{2021zndo...1011188B},
  \texttt{astropy} \citep{astropy_2013,astropy_2018,astropy_2022},
	\texttt{lightkurve} \citep{2018ascl.soft12013L},
  \texttt{numpy} \citep{2020Natur.585..357H}, 
  \texttt{pyGAM} \citep{daniel_serven_2018_1208723},
  \texttt{scipy} \citep{2020NatMe..17..261V},
  \texttt{TESS-point}  \citep{2020ascl.soft03001B},
  \texttt{wotan} \citep{2019AJ....158..143H}.
}
\ 

\facilities{
 	{\it Astrometry}:
		Gaia \citep{2018A&A...616A...1G,2023A&A...674A...1G}.
 	{\it Imaging}:
    Second Generation Digitized Sky Survey. 
 	{\it Spectroscopy}:
		Keck:I~(HIRES; \citealt{1994SPIE.2198..362V}).
 	{\it Photometry}:
 	  TESS \citep{2015JATIS...1a4003R},
  {\it Broadband photometry}:
    2MASS \citep{2006AJ....131.1163S},
    APASS \citep{2016yCat.2336....0H},
		Gaia \citep{2018A&A...616A...1G,2023A&A...674A...1G},
    SDSS \citep{2000AJ....120.1579Y},
    WISE \citep{2010AJ....140.1868W,2014yCat.2328....0C}.
}

\clearpage

\def\bibfont{\footnotesize}
\bibliographystyle{yahapj}                           
\renewcommand*{\bibfont}{\footnotesize} 
{\footnotesize \bibliography{bibliography} }


\startlongtable
\begin{deluxetable}{rrrrrrlrllrrrrrr}
\tabletypesize{\footnotesize}
\tablecaption{Bona fide, candidate, and debunked complex periodic
  variables from the TESS 2-minute data.  The non-truncated machine readable versions
  are accessible both through the online journal, and at \url{https://zenodo.org/record/8327508}. \label{tab:thetable}}
\startdata
  TIC & $T$ & $d$ & $\bprp$ & RUWE & $P$ & Assoc & Age & $T_{\rm eff}$ & $R_\star$ & $M_\star$ & $R$$_{\rm c}$ & $P_{\rm sec}$ & Quality & Bin & $N_{\rm sector}$ \\
  -- &   mag & pc &    mag &      -- &        hr &   -- &        Myr &  K                  & $R_\odot$ & $M_\odot$ & $R_\star$         & hr                     & -- & -- & -- \\
\hline
368129164 & 9.29 & 18.3 & 2.89 & 6.95 & 6.44 & ABDMG & 149 & 3127 & 0.72 & 0.4 & 1.79 & 2.60 & 1 & 011 & 3 \\
405754448 & 9.63 & 92.6 & 1.75 & 6.81 & 12.92 & LCC & 15 & 4278 & 1.5 & 0.82 & 1.74 & 134.4 & 1 & 011 & 5 \\
167664935 & 10.31 & 62.5 & 2.52 & 5.05 & 14.05 & UCL & 16 & 3316 & 1.43 & 0.38 & 1.49 & 10.71 & 1 & 011 & 3 \\
311092148 & 11.03 & 26.8 & 3.03 & 1.5 & 7.86 & COL & 42 & 3071 & 0.48 & 0.28 & 2.71 & - & 1 & 000 & 1 \\
402980664 & 11.11 & 21.3 & 3.04 & 1.48 & 18.56 & COL & 42 & 3090 & 0.37 & 0.22 & 5.78 & - & 1 & 000 & 10 \\
50745567 & 11.28 & 38.0 & 3.22 & 3.95 & 6.34 & BPMG & 24 & 2982 & 0.67 & 0.28 & 1.67 & 28.55 & 1 & 011 & 2 \\
59836633 & 11.38 & 61.9 & 2.71 & 1.21 & 14.96 & BPMG & 24 & 3243 & 0.84 & 0.45 & 2.79 & - & 1 & 000 & 3 \\
425933644 & 11.4 & 43.2 & 2.82 & 10.29 & 11.67 & THA & 45 & 3136 & 0.62 & 0.41 & 3.12 & - & 1 & 010 & 6 \\
142173958 & 11.61 & 70.6 & 3.09 & 2.9 & 11.76 & TWA & 10 & 3019 & 1.14 & 0.26 & 1.47 & 12.84 & 1 & 011 & 3 \\
146539195 & 11.62 & 48.2 & 3.37 & 2.86 & 6.73 & BPMG & 24 & 2882 & 0.81 & 0.24 & 1.37 & 7.29 & 1 & 011 & 2 \\
206544316 & 11.63 & 42.8 & 2.89 & 1.26 & 7.73 & THA & 45 & 3116 & 0.57 & 0.35 & 2.43 & - & 1 & 000 & 6 \\
335598085 & 11.9 & 105.5 & 2.85 & 2.79 & 15.85 & LCC & 15 & 3163 & 1.3 & 0.28 & 1.61 & 17.94 & 1 & 011 & 3 \\
405910546 & 12.11 & 111.9 & 2.36 & 1.09 & 37.99 & LCC & 15 & 3463 & 0.93 & 0.6 & 5.19 & - & 1 & 100 & 4 \\
272248916 & 12.15 & 80.5 & 2.83 & 5.5 & 8.9 & UCL & 16 & 3188 & 0.81 & 0.4 & 1.97 & 50.7 & 1 & 011 & 3 \\
178155030 & 12.17 & 46.8 & 2.91 & 1.29 & 11.67 & THA & 45 & 3094 & 0.49 & 0.3 & 3.53 & - & 1 & 000 & 4 \\
224283342 & 12.29 & 38.0 & 3.04 & 1.27 & 21.3 & COL & 42 & 3068 & 0.39 & 0.23 & 6.06 & - & 1 & 100 & 3 \\
89026133 & 12.31 & 131.6 & 2.82 & 4.0 & 11.2 & UCL & 16 & 3176 & 1.32 & 0.3 & 1.28 & 27.83 & 1 & 011 & 3 \\
234295610 & 12.51 & 48.1 & 3.04 & 1.13 & 18.29 & THA & 45 & 3047 & 0.45 & 0.26 & 5.0 & - & 1 & 000 & 3 \\
118449916 & 12.54 & 97.1 & 3.09 & 25.18 & 12.31 & TAU & 2 & 3022 & 1.05 & 0.28 & 1.68 & 6.71 & 1 & 011 & 4 \\
67897871 & 12.55 & 148.2 & 3.01 & 2.55 & 6.23 & USCO & 10 & 3096 & 1.5 & 0.18 & 0.65 & 6.72 & 1 & 011 & 2 \\
353730181 & 12.65 & 106.6 & 2.75 & 1.23 & 13.51 & TAU & 2 & 3221 & 0.81 & 0.4 & 2.6 & - & 1 & 000 & 4 \\
201898222 & 12.68 & 42.2 & 3.21 & 1.29 & 10.7 & THA & 45 & 2971 & 0.39 & 0.2 & 3.64 & 13.62 & 1 & 001 & 5 \\
264767454 & 12.73 & 123.3 & 2.93 & 12.3 & 10.01 & COL(?) & 42 & 3096 & 1.11 & 0.38 & 1.52 & 20.62 & 1 & 011 & 13 \\
442571495 & 12.75 & 80.8 & 3.03 & 1.64 & 9.59 & UCL & 16 & 3066 & 0.67 & 0.3 & 2.27 & 13.82 & 1 & 001 & 3 \\
2234692 & 12.8 & 53.7 & 3.0 & 1.2 & 6.52 & COL & 42 & 3077 & 0.44 & 0.25 & 2.53 & 59.8 & 1 & 001 & 7 \\
94088626 & 12.88 & 57.6 & 3.07 & 1.12 & 6.6 & ARG & 45 & 3064 & 0.46 & 0.27 & 2.51 & - & 1 & 000 & 2 \\
264599508 & 12.88 & 79.7 & 3.01 & 1.89 & 7.9 & COL & 42 & 3097 & 0.6 & 0.35 & 2.34 & 8.99 & 1 & 001 & 7 \\
363963079 & 12.92 & 83.1 & 3.09 & 8.0 & 7.82 & ARG & 45 & 3041 & 0.67 & 0.35 & 2.11 & 7.41 & 1 & 011 & 7 \\
193831684 & 13.03 & 51.6 & 3.23 & 1.16 & 31.02 & BPMG & 24 & 2979 & 0.42 & 0.2 & 6.87 & - & 1 & 000 & 3 \\
177309964 & 13.1 & 91.0 & 2.94 & 1.15 & 10.88 & CAR & 45 & 3128 & 0.61 & 0.38 & 2.94 & - & 1 & 000 & 34 \\
425937691 & 13.18 & 43.1 & 3.77 & 2.86 & 4.82 & THA & 45 & 2780 & 0.41 & 0.16 & 1.92 & 3.22 & 1 & 011 & 5 \\
141146667 & 13.28 & 57.6 & 3.28 & 1.23 & 3.93 & FIELD & NaN & 2972 & 0.42 & NaN & NaN & - & 1 & 000 & 6 \\
332517282 & 13.29 & 39.0 & 3.27 & 1.05 & 9.67 & ABDMG & 149 & 2966 & 0.28 & 0.21 & 4.84 & - & 1 & 000 & 3 \\
144486786 & 13.3 & 77.4 & 3.05 & 15.05 & 6.82 & COL & 42 & 3084 & 0.49 & 0.29 & 2.43 & 11.49 & 1 & 011 & 4 \\
38820496 & 13.3 & 44.1 & 3.37 & 1.08 & 15.73 & THA & 45 & 2916 & 0.34 & 0.16 & 5.15 & - & 1 & 000 & 5 \\
289840926 & 13.31 & 40.2 & 3.75 & 1.16 & 4.8 & BPMG & 24 & 2785 & 0.36 & 0.14 & 2.08 & 15.64 & 1 & 001 & 3 \\
404144841 & 13.33 & 77.1 & 3.19 & 1.11 & 10.74 & TWA & 10 & 3008 & 0.53 & 0.24 & 2.91 & - & 1 & 000 & 4 \\
89463560 & 13.45 & 123.9 & 2.97 & 1.31 & 9.43 & ARG & 45 & 3073 & 0.75 & 0.4 & 2.21 & 7.76 & 1 & 001 & 10 \\
300651846 & 13.49 & 109.2 & 2.86 & 1.16 & 8.26 & CAR & 45 & 3159 & 0.61 & 0.4 & 2.51 & - & 1 & 000 & 31 \\
267953787 & 13.49 & 130.5 & 3.59 & 1.2 & 17.46 & TAU & 2 & 2833 & 1.07 & 0.12 & 1.57 & - & 1 & 000 & 4 \\
68812630 & 13.6 & 123.8 & 3.22 & 1.63 & 9.04 & TAU & 2 & 3000 & 0.76 & 0.28 & 1.89 & 5.28 & 1 & 001 & 3 \\
141306513 & 13.65 & 50.2 & 3.4 & 1.08 & 13.36 & THA & 45 & 2930 & 0.32 & 0.16 & 4.79 & - & 1 & 000 & 2 \\
201789285 & 14.03 & 45.4 & 3.82 & 1.19 & 3.64 & THA & 45 & 2754 & 0.3 & 0.12 & 1.99 & - & 1 & 000 & 5 \\
294328887 & 14.23 & 97.1 & 3.22 & 1.05 & 8.51 & CARN & 200 & 2998 & 0.45 & 0.35 & 3.32 & - & 1 & 000 & 35 \\
312410638 & 14.3 & 136.9 & 3.12 & 1.09 & 28.06 & UCL & 16 & 3049 & 0.58 & 0.25 & 5.11 & - & 1 & 000 & 3 \\
38539720 & 14.52 & 129.4 & 3.37 & 1.2 & 9.16 & PERI & 120 & 2952 & 0.56 & 0.27 & 2.57 & - & 1 & 000 & 1 \\
359892714 & 14.53 & 95.5 & 4.04 & 1.07 & 11.33 & EPSC & 3 & 2669 & 0.56 & 0.13 & 2.34 & - & 1 & 000 & 6 \\
118769116 & 14.58 & 119.0 & 3.6 & 1.13 & 8.56 & TAU & 2 & 2845 & 0.55 & 0.18 & 2.18 & - & 1 & 000 & 4 \\
440725886 & 14.69 & 135.1 & 2.96 & 1.06 & 3.92 & PLE & 112 & 3117 & 0.45 & 0.35 & 1.98 & - & 1 & 000 & 5 \\
397791443 & 15.01 & 151.1 & 3.1 & 1.06 & 6.95 & IC2602 & 46 & 3023 & 0.48 & 0.26 & 2.45 & - & 1 & 000 & 6 \\
160329609 & 9.65 & 8.7 & 3.4 & 1.18 & 24.31 & ARG & 45 & 2913 & 0.35 & 0.16 & 6.62 & - & 0 & 000 & 3 \\
148646689 & 12.14 & 140.4 & 2.44 & 1.7 & 10.63 & UCL & 16 & 3441 & 1.27 & 0.55 & 1.58 & 13.37 & 0 & 001 & 3 \\
280945693 & 12.27 & 98.2 & 2.97 & 1.16 & 15.27 & LCC & 15 & 3074 & 1.09 & 0.31 & 1.94 & - & 0 & 100 & 5 \\
165184400 & 12.37 & 43.2 & 3.02 & 1.24 & 15.91 & THA & 45 & 3094 & 0.42 & 0.25 & 4.81 & - & 0 & 000 & 4 \\
245834739 & 12.55 & 115.4 & 2.85 & 1.49 & 10.47 & TAU & 2 & 3131 & 1.01 & 0.36 & 1.71 & 9.86 & 0 & 001 & 6 \\
125843782 & 13.01 & 127.7 & 2.86 & 1.21 & 44.17 & TAU & 2 & 3128 & 0.91 & 0.35 & 4.9 & - & 0 & 000 & 4 \\
244161191 & 13.17 & 44.7 & 3.54 & 1.28 & 7.17 & COL & 42 & 2867 & 0.37 & 0.16 & 2.76 & 8.39 & 0 & 001 & 3 \\
231058925 & 13.17 & 51.2 & 3.25 & 1.35 & 8.87 & THA & 45 & 2966 & 0.39 & 0.2 & 3.28 & - & 0 & 000 & 5 \\
301676454 & 13.4 & 70.7 & 3.07 & 1.24 & 9.18 & ARG & 45 & 3057 & 0.45 & 0.27 & 3.19 & - & 0 & 000 & 1 \\
58084670 & 13.58 & 140.2 & 2.82 & 1.06 & 11.16 & FIELD & NaN & 3142 & 0.77 & NaN & NaN & - & 0 & 000 & 6 \\
67745212 & 13.63 & 27.8 & 3.8 & 1.11 & 5.12 & COL & 42 & 2804 & 0.21 & 0.09 & 3.2 & - & 0 & 000 & 2 \\
5714469 & 13.73 & 78.3 & 3.65 & 1.12 & 10.35 & UCL & 16 & 2825 & 0.54 & 0.18 & 2.53 & - & 0 & 000 & 3 \\
259586708 & 13.82 & 95.6 & 2.93 & 1.17 & 22.52 & COL & 42 & 3128 & 0.46 & 0.29 & 5.81 & - & 0 & 000 & 7 \\
435903839 & 11.95 & 80.7 & 2.49 & 17.7 & 10.82 & ABDMG(?) & 149 & 3412 & 0.74 & 0.51 & 2.66 & - & -1 & 010 & 6 \\
57830249 & 11.96 & 48.8 & 3.2 & 1.34 & 43.82 & TWA & 10 & 2956 & 0.69 & 0.25 & 5.69 & - & -1 & 000 & 3 \\
193136669 & 13.06 & 61.1 & 3.49 & 1.22 & 37.64 & TWA & 10 & 2853 & 0.58 & 0.2 & 5.77 & - & -1 & 000 & 4 \\

\enddata
\tablecomments{This table includes \ngoods\ good CPVs
  (\texttt{Quality} flag 1), \nmaybes\ ambiguous CPVs
  (\texttt{Quality} flag 0), and \ndebunked\ impostors
  (\texttt{Quality} flag -1).  The three-bit binarity flag ``Bin'' is
  for Gaia DR3 \texttt{radial\_velocity\_error} outliers (bit 1), Gaia
  DR3 \texttt{ruwe} outliers (bit 2), and stars with multiple TESS
  periods (bit 3).  The machine-readable version, available online,
  includes additional columns for the Gaia DR2 and DR3 source
  identifiers, as well as the stellar parameter uncertainties.  The
  age uncertainties are typically $\approx \pm 10\%$, but can be
  asymmetric.  The median statistical uncertainties on the
  temperature, radius, and mass are $\pm$50\,K, $\pm$4\% and $\pm$9\%
  respectively.  $N_{\rm sector}$ denotes the number of TESS sectors
  for which {\it any} data are expected to be acquired between
  2018~July and 2024~Oct.  This number is generally greater than the
  number of sectors for which 2-minute cadence data exist.
  Association names and provenance follow conventions adopted by
  \citet{2018ApJ...856...23G}: 
	ABDMG: AB~Doradus moving group \citep{2015MNRAS.454..593B}.
	ARG: Argus \citep{2019ApJ...870...27Z}.
	BPMG: $\beta$~Pic moving group \citep{2015MNRAS.454..593B}.
  CARN: Carina Near moving group \citep{2006ApJ...649L.115Z}.
	COL: Columba \citep{2015MNRAS.454..593B}.
	EPSC: $\epsilon$ Chamaeleontis \citep{2013MNRAS.435.1325M}.
	LCC: Lower Centaurus Crux \citep{2016MNRAS.461..794P}.
	PERI: Pisces-Eridani \citep{2019AJ....158...77C}.
	PLE: Pleiades \citep{2015ApJ...813..108D}.
	TAU: Taurus \citep{1995ApJS..101..117K}.
	THA: Tucana-Horologium assocation \citep{2015MNRAS.454..593B}.
	TWA: TW Hydrae assocation \citep{2015MNRAS.454..593B}.
	UCL: Upper Centaurus Lupus \citep{2016MNRAS.461..794P}.
	USCO: Upper Scorpius \citep{2016MNRAS.461..794P}.
	The ``(?)'' string denotes low-confidence membership.
	}
\end{deluxetable}


\appendix

\section{The ring hypothesis}
\label{app:ring}

One hypothesis for the CPVs, presented by \citet{2019ApJ...876..127Z},
is that the star might be ``{\it orbited by one or more rings composed
of dust-size or somewhat larger particles\ldots The ring particles
would move in Keplerian orbits at relatively large distances from the
star, and therefore the sublimation lifetime would not be an issue
even if the particles are dust-like in size.}'' A sketch of this
scenario was presented by \citet{2019ApJ...876..127Z}, in their
Figure~11.  An example set of proposed parameters involved a ring
inclined with respect to the stellar spin axis by a few degrees, and
with inner and outer radii of 10 and 15 stellar radii.

One concern with the ring hypothesis is that if a cool spot were to
transit behind the ring, it would produce a brightening, not a
dimming.   Most CPVs show dimmings.   The ring scenario would therefore
imply that large hot spots are common in the photospheres of
pre-main-sequence M dwarfs.  Empirical evidence however suggests that
cool spots dominate the optical variability of disk-free
pre-main-sequence stars.  This evidence includes flux excursions
caused by spot-crossings during planetary transits
\citep[e.g.][]{2020AJ....160...33R,2022AJ....163..147G}, correlations
between simultaneous photometric and chromospheric time-series
\citep{2019A&A...621A..21R}, and stellar spectra that show molecules
that only form at cool temperatures
\citep[e.g.][]{2017ApJ...836..200G,2023ApJ...946...10P}.

An independent  concern with the ring hypothesis is that it is
fine-tuned.  The model requires specific locations for the inner edge
and the outer edge of the ring, an inclination that yields a band with
a specific apparent size, and material in the ring that must be
optically thick while also being homogeneous enough to not induce any
apparent photometric variability.  It is challenging to ascribe
specific probabilities to any one of these factors.  However the
requirement that they all be simultaneously met seems sufficiently
severe to disfavor this scenario.

\begin{figure*}[!t]
	\begin{center}
		\subfloat{
			\includegraphics[width=0.48\textwidth]{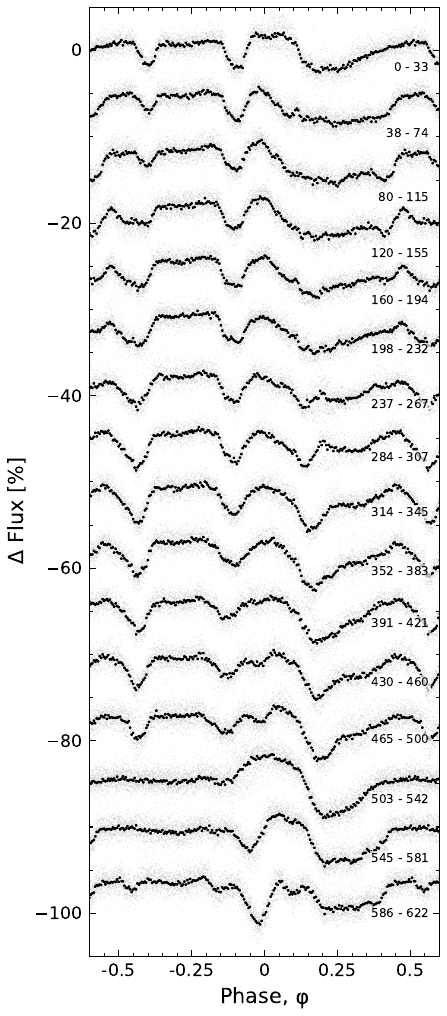}
			\includegraphics[width=0.48\textwidth]{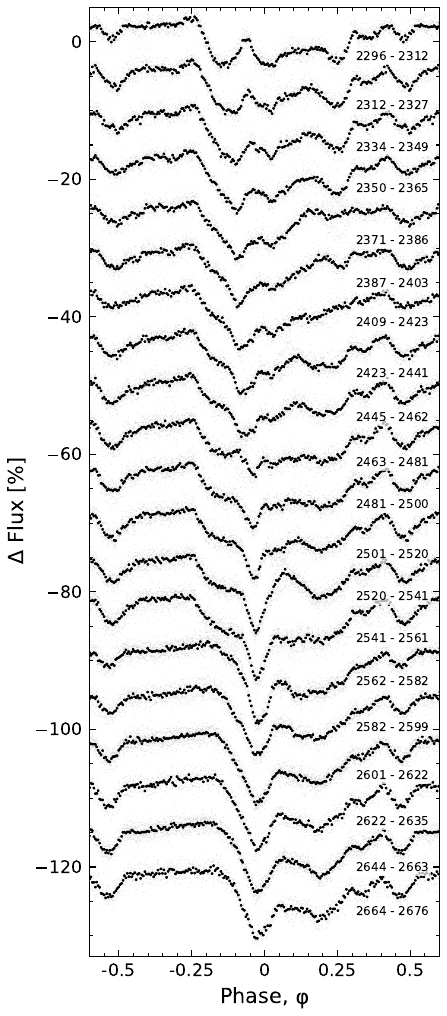}
		}
	\end{center}
	\vspace{-0.4cm}
	\caption{
		{\bf Light curve evolution of TIC 300651846}.
		All available 2-minute cadence data as of 2023 Aug 11 are shown.
		Cycles 0 to 622 span TESS Sectors 32-39 (Nov 2020--June 2021);
		cycles 2296-2676 span Sectors 61-65 (Jan--June 2023).  We assumed
		a 8.254\,hr period and a fixed reference epoch (BTJD 2174.127) for
		both panels.  Light curve segments are split based on the presence
		of gaps longer than three hours.  Cycle numbers are listed in the
		lower-right of each light curve segment.
	}
	\label{fig:tic3006timegroups}
\end{figure*}

\begin{figure*}[!t]
	\begin{center}
		\subfloat{
			\includegraphics[width=0.48\textwidth]{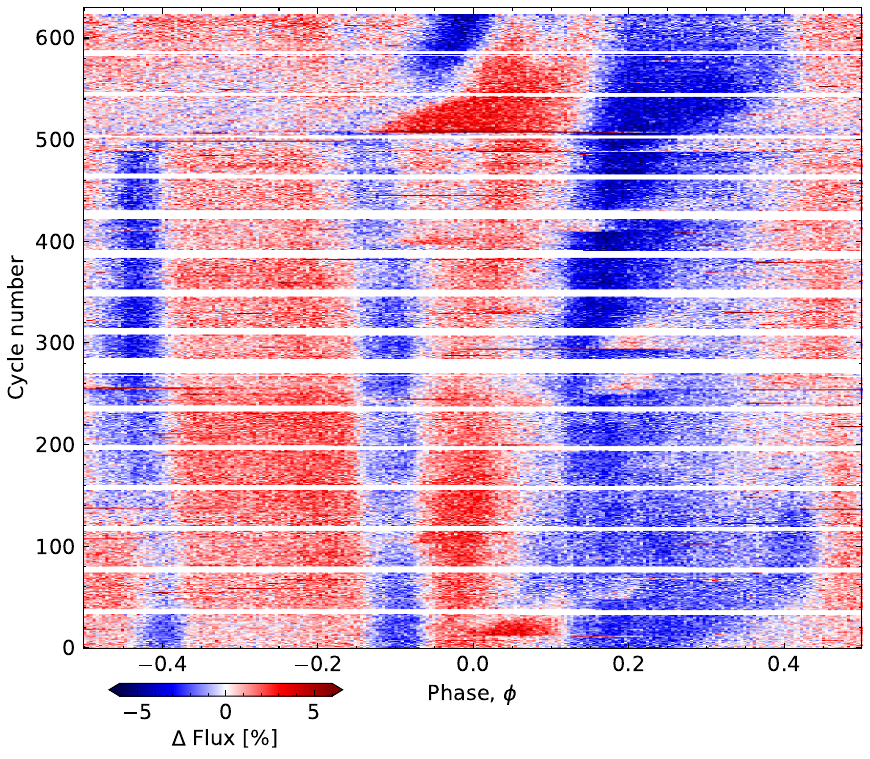}
			\includegraphics[width=0.48\textwidth]{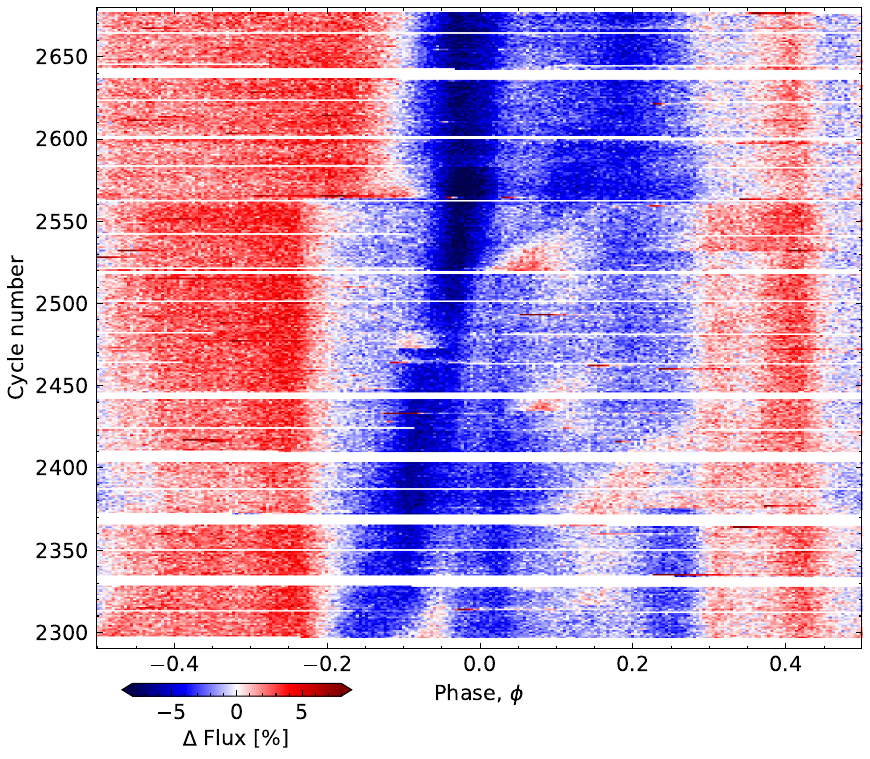}
		}
	\end{center}
	\vspace{-0.4cm}
	\caption{
		{\bf River plots of TIC 300651846}.
		This is an alternative visualization of the data in
		Figure~\ref{fig:tic3006timegroups}.  All available 2-minute
		cadence data as of 2023 Aug 11 are shown.  Cycles 0 to 622 span
		TESS Sectors 32-39 (Nov 2020--June 2021); cycles 2296-2676 span
		Sectors 61-65 (Jan--June 2023).  We assumed $P$$=$8.254\,hr and
		$t_0$=2174.127 [BTJD].  Note that the two panels have slightly
		different color scales.
	}
	\label{fig:tic3006river}
\end{figure*}

\section{TIC~300651846}
\label{app:tic3006}

Figures~\ref{fig:tic3006timegroups} and~\ref{fig:tic3006river} show
2-minute cadence data for TIC~300651846, a CPV in the TESS
continuous viewing zone.  If it were not for the existence of
LP~12-502, this source would have received greater
attention.  With the exception of a few sectors, TESS data will exist
for TIC~300651846 for at least Sectors 1-12, 27-39, and 61-69.  While
most of the available data exist in the full frame images,
Figures~\ref{fig:tic3006timegroups} and~\ref{fig:tic3006river} focus
only on the currently available 2-minute cadence data.

During Sectors 32-39, the source shows between one and four local
minima per cycle.  During the early portions of Sectors 61-65, it is
more complex, with at least five clear local minima per
cycle.  As the source evolves, its shape becomes simpler, and the
sharpness of one global minimum appears to increase.

State-switches analogous to those observed in LP 12-502 occur at
cycles 498 and 2554.  During the cycle 498 switch, two narrow dips at
$\phi$$\approx$$-0.4$ and $\phi$$\approx$0.0 collapse.  For the cycle
2554 switch, a longer dip collapses.  This is visible as a change in
curvature in Figure~\ref{fig:tic3006timegroups} between $\phi \in
[-0.25, -0.05]$ across cycles 2520-2561.  The  more typical dip
evolution timescales for TIC~300651846 seem to be $\approx$50-100
cycles.  Unlike the LP~12-502 river plots
(Figure~\ref{fig:lpriver0}), we did not subtract any ``continuum
sinusoid'' for this source, because the continuum is not as obviously
defined.

\section{Literature comparison}
\label{sec:litcomp}

We compared our CPV search results against previous work by
compiling a list of CPVs from both K2
\citep{2017AJ....153..152S,2018AJ....155...63S} and TESS
\citep{2019ApJ...876..127Z,2020AJ....160...86B,2021AJ....161...60S,2022AJ....163..144G,2023ApJ...945..114P}.
We counted PTFO 8-8695 as a TESS-detected CPV \citep[but see][and
references therein]{2020AJ....160...86B}.  This effort resulted in a
list of 74 unique objects that had been reported to be CPVs.  We made
no attempt to reclassify these sources based on our subjective grading
scheme.  TESS contributed 41 of the 74 objects; the remaining 33 were
from K2.  

{\it How many known CPVs did we miss?}---A minority of the literature
CPVs, 19/74, were stars with 2-minute TESS data acquired between
Sectors 1-55 that also met conditions~\ref{eq:one}--\ref{eq:four}.
Our blind search (Table~\ref{tab:thetable}) recovered all but two of
these 19 sources: TIC~65347864 in Tuc-Hor \citep{2023ApJ...945..114P},
and TIC~243499565 in Sco-Cen \citep{2021AJ....161...60S}.  During our
blind vetting, we manually labelled both sources as eclipsing
binaries.  Reconsidering with the knowledge of previous literature, we
believe that TIC~65347864 remains ambiguous: this source shows one
persistent broad local minimum superposed over sinusoidal spot
modulation; it could be an eclipsing binary, an RS CVn, or a CPV.

TIC~243499565 however is a bona fide CPV; our blind labeling for this
source was incorrect.  \citet{2021AJ....161...60S} made the
classification based on Sector~11 full-frame image data.  During
Sector~38, the dip phases were different, and the source
resembled an eccentric eclipsing binary (similar to TIC~193831684).  Sector~64 recently showed
the development of additional local minima in the light curve so that
it shows three dips per cycle; its CPV classification is secure.  Our
search therefore missed one star that had previously been classified
as a CPV.

{\it How many of our CPVs are new?}---Considering only our ``good''
CPVs, 35 of these 50 sources have not yet been reported in the
literature.  Considering only the 13 ``ambiguous'' CPVs, 11 are new to
the literature.  Two sources, TIC~67897871 (EPIC~202724025; RIK-90;
\citealt{2017AJ....153..152S}) and TIC~118769116 (EPIC~247343526;
\citealt{2017AJ....153..152S}), were previously only known to be CPVs
due to K2 observations.  Both periods appear to have remained
identical, within measurement precision.  The shapes have of course changed.
These two objects, combined with PTFO 8-8695, provide evidence
that CPV variability persists over timescales of at least decades.

\begin{figure}[!t]
	\begin{center}
		\centering
		\includegraphics[width=0.55\textwidth]{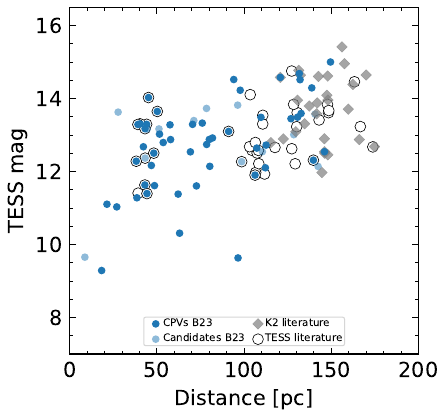}
		\vspace{-0.2cm}
		\caption{
			{\bf Brightnesses and distances of our CPVs compared
      against previous work.}  
      ``K2 literature'' corresponds to \citet{2017AJ....153..152S} and
      \citet{2018AJ....155...63S}.  ``TESS literature'' corresponds to
      \citet{2019ApJ...876..127Z,2021AJ....161...60S,2022AJ....163..144G}, and \citet{2023ApJ...945..114P}.
      ``\replaced{CQVs}{CPVs}'' and ``candidates'' are as in
      Figure~\ref{fig:catalogscatter}, though for visual clarity
      unresolved binaries are not highlighted.  PTFO 8-8695, a CPV at
      a distance of $\approx$350\,pc, is not shown.
		}
			\vspace{-0.5cm}
		\label{fig:tmagvsdist}
	\end{center}
\end{figure}

{\it How do the brightness and distance distributions
compare?}---Figure~\ref{fig:tmagvsdist} compares the TESS magnitudes
and distances for the CPVs in Table~\ref{tab:thetable} against the
previously known literature CPVs.  At the extremes, our new sources
include the closest and brightest CPVs currently known. 
All K2 CPVs are in Sco-Cen,
$\approx$100-180\,pc from the Sun.  The large gap in the TESS literature CPV
distance distribution was because the studies by
\citep{2019ApJ...876..127Z} and \citet{2023ApJ...945..114P} both
favored Tuc-Hor; many of our new \replaced{CQVs}{CPVs} are within this 40-100\,pc gap.

\section{No significant power at 20~second cadence}
\label{sec:twentysec}

TESS was the first instrument to show that CPV light curves contain
power at timescales of a few minutes
\citep{2019ApJ...876..127Z,2022AJ....163..144G}.  This advance was
enabled by the fifteen-fold faster cadence in the TESS 2-minute data,
relative to K2.  A logical follow-up is to ask whether the periodic
components of the CPV light curves contain power at timescales below
one minute.  Between 2020 and 2021, we observed 10 CPVs at 20-second
cadence with TESS in order to explore this question (TESS DDT029; PI
L.~Bouma).  The stars were TICs 142173958, 146539195, 24518895,
276453848, 264599508, 363963079, 144486786, 408188366, 300651846,
262400835.  These sources were selected from CPVs known at the time to
have short periods and sharp features when observed at 2-minute
cadence.  Comparing the 20-second to 2-minute data for these stars
(data available on MAST), we concluded that these CPVs did not contain
appreciable power at timescales shorter than a few minutes, other than
the usual flaring.  This is consistent with the expected few-minute
ingress and egress timescales for transiting circumstellar material.

\section{The CPVs are not obviously accreting}
\label{sec:accretion}

\begin{figure*}[!t]
	\begin{center}
	\subfloat{
			\includegraphics[width=0.7\textwidth]{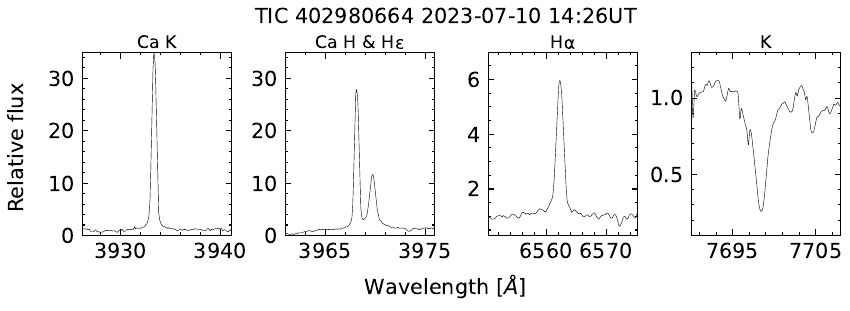}
    }

	\vspace{-1.cm}
	\subfloat{
			\includegraphics[width=0.7\textwidth]{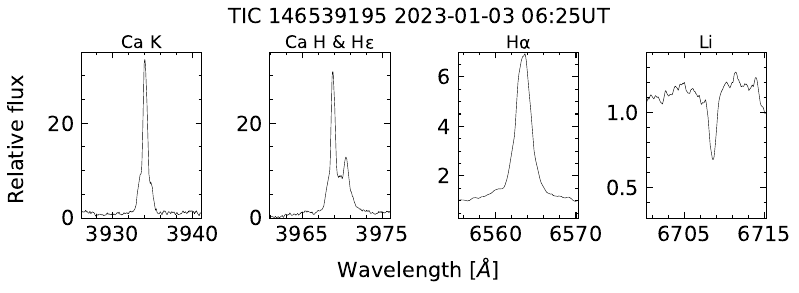}
    }

	\vspace{-1.cm}
	\subfloat{
			\includegraphics[width=0.7\textwidth]{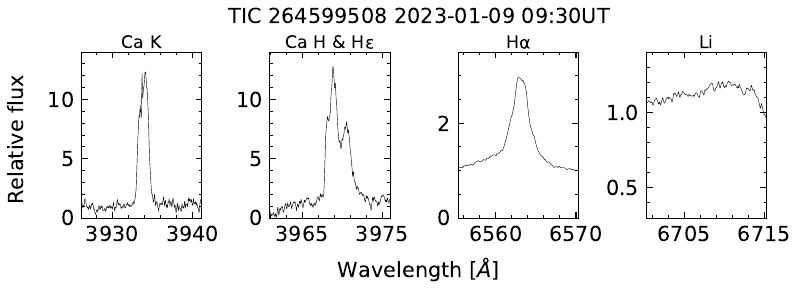}
    }
	\end{center}
	\vspace{-0.4cm}
	\caption{
		{\bf Spectral age and activity diagnostics for three CPVs}.
    Wavelengths are in air; the continuum normalization is relative to
    the entire order.  The H$\alpha$ emission strength classifies the
    stars as weak-lined T Tauris.  The lithium detection for
    TIC~146539195 is consistent with its mass and $\beta$~Pic
    membership; the non-detections for LP~12-502 (TIC~402980664) and
    TIC~264599508 are consistent with the $\approx$42\,Myr age implied
    by their membership in the Columba moving group.
	}
	\label{fig:speccutouts}
\end{figure*}

We acquired iodine-free reconnaissance spectra using Keck/HIRES for
three CPVs.  The goals were to determine the chromospheric activity
levels, and to check for indications of either accretion or
spectroscopic binarity.  We acquired a 15 minute exposure of
TIC~146539195 on 2023 January 3, a 15 minute exposure of TIC~264599508
on 2023 January 9, and a 30 minute exposure of LP~12-502
(TIC~402980664) on 2023 July 10.  The acquisition and analysis
followed the usual techniques of the California Planet Survey
\citep{2010ApJ...721.1467H}.  Figure~\ref{fig:speccutouts} shows
cutouts from the resulting spectra, centered on the \ion{Ca}{2} HK
windows, H$\alpha$, and the \ion{Li}{1} 6708\,\AA\ doublet.  The
\ion{Ca}{2} H emission line is blended with H$\epsilon$.  While a more
detailed analysis will be left for future work, these spectra confirm
previous understanding established by \citet{2017AJ....153..152S} that
the stars are chromospherically active M dwarfs in the ``weak-lined''
T Tauri regime \citep[e.g.][Figure~15]{2019AJ....157...85B}.  Their
H$\alpha$ equivalent widths, at $\approx$14\,\AA, $\approx$3\AA, and
$\approx$8\,\AA\ (for TIC~264599508, 146539195, and 402980664
respectively) are consistent with purely chromospheric emission.  The
blue excess in TIC~264599508 could be explained by a second unresolved
star; the TESS light curve for this source shows both the 7.90\,hr CPV
signal, and a 9.00\,hr rotation signal with comparable amplitude.

\clearpage
\listofchanges

\end{document}